\documentclass{aa}
\pdfoutput=1

\usepackage[colorlinks=true,citecolor=blue,pagebackref=false,pdftex]{hyperref}
\usepackage[nointegrals]{wasysym}                 
\usepackage{graphicx,subfig}
\usepackage{amsmath,amssymb,amscd,verbatim,xkeyval,bm,array,bigints}
\usepackage{enumitem,booktabs,textcomp}
\usepackage{epstopdf}

\usepackage{rotating}

\usepackage{makecell}
\usepackage[table,x11names]{xcolor}
\usepackage{tikz}

\usepackage{natbib}
\bibpunct{(}{)}{;}{a}{}{,}

\usepackage{changes}

\newcommand\eg{{\rm e.g.} }

\author{Mara Volpi\inst{1}, Arnaud
  Roisin\inst{1} and Anne-Sophie Libert\inst{1}}
 
\begin{document}

\title{On the 3D secular dynamics of radial-velocity-detected planetary systems}
\titlerunning{3D secular dynamics of RV-detected planetary systems}
\institute{naXys, Department of Mathematics, University of Namur, Rempart de la Vierge 8, B-5000 Namur, Belgium}

\date{}

\abstract {} {To date, more than 600 multi-planetary systems have been
  discovered. Due to the limitations of the detection methods, our
  knowledge of the systems is usually far from complete. In
  particular, for planetary systems discovered with the radial
  velocity (RV) technique, the inclinations of the orbital planes, and
  thus the mutual inclinations and planetary masses, are unknown. Our
  work aims to constrain the spatial configuration of several
  RV-detected extrasolar systems that are not in a mean-motion
  resonance.}  {Through an analytical study based on a first-order
  secular Hamiltonian expansion and numerical explorations performed
  with a chaos detector, we identified ranges of values for the
  orbital inclinations and the mutual inclinations, which ensure the
  long-term stability of the system. Our results were validated by
  comparison with n-body simulations, showing the accuracy of our
  analytical approach up to high mutual inclinations ($\sim
  70^\circ$-$80^\circ$).}  {We find that, given the current
  estimations for the parameters of the selected systems, long-term
  regular evolution of the spatial configurations is observed, for all
  the systems, i) at low mutual inclinations (typically less than
  $35^\circ$) and ii) at higher mutual inclinations, preferentially if
  the system is in a Lidov-Kozai resonance. Indeed, a rapid
  destabilisation of highly mutually inclined orbits is commonly
  observed, due to the significant chaos that develops around the
  stability islands of the Lidov-Kozai resonance. The extent of the
  Lidov-Kozai resonant region is discussed for ten planetary systems
  (HD~$11506$, HD~$12661$, HD~$134987$, HD~$142$, HD~$154857$,
  HD~$164922$, HD~$169830$, HD~$207832$, HD~$4732$, and HD~$74156$).}
          {} 

\maketitle

\begin{table*}
\centering
\caption{Orbital parameters of the selected systems.}
\label{tab:orbpar}
\begin{tabular}{llllllll}
\hline\hline System & & $m\sin i\,(M_J)$ & $M_{\text{Star}}\,
(M_{\odot})$ & a & e & $\omega$ & References\\ \hline HD 11506 & b &
3.44 & 1.19 & 2.43 & 0.22 & 257.8 & \citet{tuo-kot-AA-2009}\\ & c &
0.82 & & 0.639 & 0.42 & 234.9 & \\ HD 12661 & b & 2.3 {\small ($\pm
  0.19$)} & 1.07 & 0.831 {\small ($\pm 0.048$)} & 0.378 {\small ($\pm
  0.0077$)} & 296 {\small ($\pm 1.5$)} &
\citet{wri-et-al-APJ-2009}\\ & c & 1.57 {\small ($\pm 0.07$)} & & 2.56
      {\small ($\pm 0.17$)} & 0.031 {\small ($\pm 0.022$)} & 165
      {\small ($\pm 0.0$)} & \\ HD 134987 & b & 1.59 & 1.07 & 0.81 &
      0.233 & 252.7 & \citet{jon-et-al-MNRAS-2010}\\ & c & 0.82 & &
      5.8 & 0.12 & 195 & \\ HD 142 & b & 1.25 {\small ($\pm 0.15$)} &
      1.1 & 1.02 {\small ($\pm 0.03$)} & 0.17 {\small ($\pm 0.06$)} &
      327 {\small ($\pm 26$)} & \citet{wit-et-al-APJ-2012}\\ & c & 5.3
      {\small ($\pm 0.7$)} & & 6.8 {\small ($\pm 0.5$)} & 0.21 {\small
        ($\pm 0.07$)} & 250 {\small ($\pm 20$)} & \\ HD 154857 & b &
      2.24 & 1.718 & 1.291 & 0.46 & 57 &
      \citet{wit-et-al-APJ-2014}\\ & c & 2.58 & & 5.36 & 0.06 & 32 &
      \\ HD 164922 & b & 0.3385 & 0.874 & 2.115 & 0.126 & 129 &
      \citet{ful-et-al-APJ-2016}\\ & c & 0.0406 & & 0.3351 & 0.22 & 81
      & \\ HD 169830 & b & 2.88 & 1.4 & 0.81 & 0.31 & 148 &
      \citet{may-et-al-AA-2004}\\ & c & 4.04 & & 3.6 & 0.33 & 252 &
      \\ HD 207832 & b & 0.56 & 0.94 & 0.57 & 0.13 & 130.8 &
      \citet{hagh-et-al-APJ-2012}\\ & c & 0.73 & & 2.112 & 0.27 &
      121.6 & \\ HD 4732 & b & 2.37 & 1.74 & 1.19 & 0.13 & 35 &
      \citet{sato-et-al-APJ-2013}\\ & c & 2.37 & & 4.6 & 0.23 & 118 &
      \\ HD 74156 & b & 1.778 & 1.24 & 0.2916 & 0.638 & 175.35 &
      \citet{feng-et-al-APJ-2015}\\ & c & 7.997 & & 3.82 & 0.3829 &
      268.9 & \\ \hline
\end{tabular}
\end{table*}

\section{Introduction}
\label{sec:intro}
The number of detected multi-planetary systems continually
increases. Despite the rising number of discoveries, our knowledge of
the physical and orbital parameters of the systems is still partial
due to the limitations of the observational techniques. Nevertheless,
it is important to acquire a deeper knowledge of the detected
extrasolar systems, in particular a more accurate understanding of the
architecture of the systems. Regarding the two-planet systems detected
via the radial velocity (RV) method, we have fairly precise data about
the planetary mass ratio, the semi-major axes, and the
eccentricities. However, we have no information either on the orbital
inclinations $i$ (i.e.  the angles the planetary orbits form with the
plane of the sky) -- which means that only minimal planetary masses
can be inferred -- or on the mutual inclination between the planetary
orbital planes $i_{mut}$. This raises questions about possible
three-dimensional (3D) configurations of the detected planetary
systems. Let us note that a relevant clue to the possible existence of
3D systems has been provided for $\upsilon$ Andromedae $c$ and $d$,
whose mutual inclination between the orbital planes is estimated to be
$30^\circ$ \citep{deit-et-al-ApJ-2015}.

A few studies on the dynamics of extrasolar systems have been devoted
to the 3D problem. Analytical works by
\citet{mich-ferr-beau-ICARUS-2006}, \citet{lib-hen-icarus-2007}, and
\citet{lib-hen-CeMDA-2008} investigated the secular evolution of 3D
exosystems that are not in a mean-motion resonance. They showed that
mutually inclined planetary systems can be long-term stable. In
particular, these works focused on the analysis of the equilibria of
the 3D planetary three-body problem, showing the generation of stable
Lidov-Kozai (LK) equilibria \citep{Lidov1962,kozai-1962} through
bifurcation from a central equilibrium, which itself becomes unstable
at high mutual inclination. Thus, around the stability islands of the
LK resonance, which offers a secular phase-protection mechanism and
ensures the stability of the system, chaotic motion of the planets
occurs, limiting the possible 3D configurations of planetary systems.

Using n-body simulations, \citet{lib-tsi-AA-2009} investigated the
possibility that five extrasolar two-planet systems, namely $\upsilon$
Andromedae, HD 12661, HD 169830, HD 74156, and HD 155358, are actually
in a LK-resonant state for mutual inclinations in the range
$[40^{\circ},60^{\circ}]$. They showed that the physical and orbital
parameters of four of the systems are consistent with a LK-type
orbital motion, at some specific values of the mutual inclination,
while around $30\%-50\%$ of the simulations generally lead to chaotic
motion. The work also suggests that the extent of the LK-resonant
region varies significantly for each planetary system considered.

Extensive long-term n-body integrations of five hierarchical
multi-planetary systems (HD 11964, HD 38529, HD 108874, HD 168443, and
HD 190360) were performed by \cite{veras-ford-ApJ-2010}. They showed a
wide variety of dynamical behaviour when assuming different
inclinations of the orbital plane with respect to the line of sight
and mutual inclinations between the orbital planes. They often reported
LK oscillations for stable highly inclined systems.

In \citet{dawson-chiang-Scie-2014}, the authors presented evidence
that several eccentric warm Jupiters discovered with eccentric giant
companions are highly mutually inclined (i.e. with a mutual
inclination in the range $[35^{\circ},65^{\circ}]$). For instance, this
is the case of the HD~$169830$ and HD~$74156$ systems, which will also be
discussed in the present work.

Recently, \cite{vol-et-al-cemda-2018} used a reverse KAM method
\citep{kolmogorov-1954,arnold-RuMaS-1963,moser-Mat-1962} to estimate
the mutual inclinations of several low-eccentric RV-detected
extrasolar systems (HD 141399, HD 143761, and HD 40307). This
analytical work addressed the long-term stability of planetary systems
in a KAM sense, requiring that the algorithm constructing KAM
invariant tori is convergent. This demanding condition leads to upper
values of the mutual inclinations of the systems close to $\sim
15^{\circ}$.

In the spirit of \cite{lib-tsi-AA-2009}, the aim of the present work
is to determine the possible 3D architectures of RV-detected systems
by identifying ranges of values for the mutual inclinations that
ensure the long-term stability of the systems. Particular attention
will be given to the possibility of the detected extrasolar systems
being in a LK-resonant state, since it offers a secular
phase-protection mechanism for mutually inclined systems, even though
the two orbits may suffer large variations both in eccentricity and
inclination. Indeed, the variations occur in a coherent way, such that
close approaches do not occur and the system remains stable.

To reduce the number of unknown parameters to take into account, we
use an analytical approach, expanding the Hamiltonian of the
three-body problem in power series of the eccentricities and
inclinations. Being interested in the long-term stability of the
system, we consider its secular evolution, averaging the Hamiltonian
over the fast angles. Thanks to the adoption of the Laplace plane, we
can further reduce the expansion to two degrees of freedom. It was
shown in previous works \citep[see for example
][]{lib-hen-icarus-2007,lib-sans-CeMDA-2013} that if the planetary
system is far from a mean-motion resonance, the secular approximation
at the first order in the masses is accurate enough to describe the
evolution of the system. Such an analytical approach is of interest for
the present purpose, since, being faster than pure n-body simulations
which also consider small-period effects, it allows us to perform an
extensive parametric exploration at a reasonable computational
cost. Moreover, in the present work we will show that the analytical
expansion is highly reliable, fulfilling its task up to high values of
the mutual inclination.

The goal of the present work is twofold. On the one hand, we study the
3D secular dynamics of ten RV-detected extrasolar systems, identifying
for each one the values in the parameter space $(i_{mut},i)$ that induce
a LK-resonant behaviour of the system. On the other hand, through
numerical explorations performed with a chaos detector, we identify
the ranges of values for which a long-term stability of the orbits is
observed, unveiling for each system the extent of the chaotic region
around the LK stability islands.

The paper is organised as follows. In Sect.~\ref{sec:model}, we
describe the analytical secular approximation and discuss its accuracy
to study the 3D dynamics of planetary systems in
Sect. ~\ref{sec:parametric}, as well as the methodology of our
parametric study. The question of possible 3D configurations of
RV-detected planetary systems is addressed in
Sect.~\ref{sec:results}. Our results are finally summarised in
Sect.~\ref{sec:ccl}.

\begin{table*}[htpb]
\centering
\caption{Convergence {\it au sens des astronomes} for the ten
  systems. The value $H_j$ corresponds to the sum of all the terms of
  the Hamiltonian given by Eq.~\eqref{eq:sec_ham} of order $j$ in
  eccentricities and inclinations. The last column gives the relative
  error between the secular Hamiltonian computed by numerical
  quadrature and the expansion of Eq.~(\ref{eq:sec_ham}). The values
  are computed for the initial condition
  $(i_{mut},i)=(50^\circ,50^\circ)$.}
\label{tab:conv}
\begin{tabular}{lcccccccc}
\hline\hline System & $H_2$ & $H_4$ & $H_6$ & $H_8$ & $H_{10}$ &
$H_{12}$ & $H_{12}/H_2$ & Relative error\\ 
\hline 
HD 11506 & 9.47e-06 & 1.66e-05 & 6.31e-06 & 7.15e-06 & 2.46e-06 & 4.73e-07 & $\mathcal{O}(10^{-2})$ & 1.62e-04\\ 
HD 12661 & 3.76e-05 & 7.24e-05 & 7.33e-05 & 3.96e-05 & 9.69e-06 & 9.25e-07 & $\mathcal{O}(10^{-2})$ & 1.42e-04\\ 
HD 134987 & 4.42e-07 & 2.61e-08 & 1.53e-09 & 1.90e-10 & 1.09e-11 & 1.62e-12 & $\mathcal{O}(10^{-6})$ & 1.10e-05\\ 
HD 142 & 4.27e-06 & 1.67e-06 & 1.40e-07 & 1.93e-08 & 1.58e-09 & 5.53e-10 & $\mathcal{O}(10^{-4})$ & 1.34e-04\\ 
HD 154857 & 2.67e-05 & 2.23e-07 & 1.20e-07 & 7.54e-08 & 1.36e-08 & 2.10e-09 & $\mathcal{O}(10^{-4})$ & 9.33e-05\\ 
HD 164922 & 1.24e-07 & 9.67e-09 & 4.53e-10 & 2.99e-11 & 5.95e-13 & 1.41e-13 & $\mathcal{O}(10^{-6})$ & 1.03e-05\\ 
HD 169830 & 7.01e-06 & 2.40e-05 & 6.51e-09 & 7.36e-06 & 2.75e-06 & 6.26e-07 & $\mathcal{O}(10^{-1})$ & 1.46e-04\\ 
HD 207832 & 9.15e-07 & 6.85e-07 & 2.08e-07 & 1.79e-07 & 4.94e-08 & 6.47e-09 & $\mathcal{O}(10^{-2})$ & 5.28e-05\\ 
HD 4732 & 6.22e-06 & 1.35e-06 & 7.42e-08 & 8.36e-08 & 1.48e-08 & 1.44e-09 & $\mathcal{O}(10^{-4})$ & 7.58e-05\\ 
HD 74156 & 2.93e-05 & 1.65e-05 & 4.68e-06 & 5.36e-05 & 3.24e-05 & 1.62e-05 & $\mathcal{O}(10^{-1})$ & 1.01e-04\\
\hline
\end{tabular}
\end{table*}

\section{Analytical secular approximation}
\label{sec:model}

We focus on the three-body problem of two exoplanets revolving around
a central star.  The indexes $0$, $1,$ and $2$ refer to the star, the
inner planet, and the outer planet, respectively. Since the total
angular momentum vector $\mathbf{C}$ is an integral of motion of the
problem, we adopt as a reference plane the constant plane orthogonal
to $\mathbf{C}$, the so-called Laplace plane. In this plane, the
Hamiltonian formulation of the problem no longer depends on the
two angles $\Omega_1$ and $\Omega_2$, but only on their constant
difference $\Omega_1-\Omega_2=\pi$. Thus, thanks to the reduction of
the nodes, the problem is reduced to four degrees of freedom. We adopt
the Poincar\'e variables,
\begin{equation}
\begin{aligned}
\Lambda_i &= \beta_i \sqrt{\mu_i a_i}, &\xi_i &= \sqrt{2\Lambda_i}
\sqrt{1-\sqrt{1-e_i^2}} \cos{\omega_i},\\ \lambda_i &= M_i+\omega_i,
&\eta_i &= -\sqrt{2\Lambda_i} \sqrt{1-\sqrt{1-e_i^2}} \sin{\omega_i},
\end{aligned}
\end{equation}
where $a$, $e$, $\omega$, and $M$ refer to the semi-major axis,
eccentricity, argument of the pericenter, and mean anomaly,
respectively, and with
\begin{equation}
  \mu_i = \mathcal{G} (m_0 + m_i), \quad \beta_i =\frac{m_0m_i}{m_0+m_i},
\end{equation}
for $i=1,2\,$. Moreover, we consider the parameter $D_2$ \citep[as
  defined in][]{rob-1995}
\begin{equation}
  \label{eq:d2}
  D_2 = \frac{(\Lambda_1 + \Lambda_2)^2 - C^2}{\Lambda_1\Lambda_2}\,,
\end{equation}
which measures the difference between the actual norm of the total
angular momentum vector $\mathbf{C}$ and the one the system would have
if the orbits were circular and coplanar (by definition, $D_2$ is
quadratic in eccentricities and inclinations).

We introduce the translation $L_j=\Lambda_j - \Lambda^*_j$, where
$\Lambda^*_j$ is the value of $\Lambda_j$ for the observed
semi-major axis $a_j$, for $j=1,2$.  We then expand the Hamiltonian in
power series of the variables ${\bm L}$, ${\bm \xi}$, ${\bm \eta,}$
and the parameter $D_2$ and in Fourier series of ${\bm \lambda}$, as
in \cite{vol-et-al-cemda-2018},
\begin{equation}
\begin{aligned}
  \label{eq:ham_comp}
  H(D_2,{\bm L},{\bm \lambda},{\bm \xi},{\bm \eta})= &
  \sum_{j_1=1}^{\infty}h^{({\rm Kep})}_{j_1,0}({\bm L})\\ &+
  \sum_{s=0}^{\infty}\sum_{j_1=0}^{\infty}\sum_{j_2=0}^{\infty}D_2^s\,%
  h_{s;j_1,j_2}({\bm L},{\bm \lambda},{\bm \xi},{\bm \eta})\,,
\end{aligned}
\end{equation}
where
\begin{itemize}
\item $h^{({\rm Kep})}_{j_1,0}$ is a homogeneous polynomial function
  of degree $j_1$ in ${\bm L}$;
\item $h_{s;j_1,j_2}$ is a homogeneous polynomial function of degree
  $j_1$ in ${\bm L}$, degree $j_2$ in ${\bm \xi}$ and ${\bm \eta}$,
  and with coefficients that are trigonometric polynomials in ${\bm
    \lambda}$.
\end{itemize}

As we are interested in the secular evolution of the system, the
Hamiltonian can be averaged over the fast angles, 
\begin{equation}
\label{eq:sec_ham}
  H(D_2,{\bm \xi},{\bm \eta}) = \sum_{j=0}^{{ORDECC}/2}
  \mathcal{C}_{j,{\bm m},{\bm n}}\,D_2^j \sum_{{\bm m}+{\bm
  n}=0}^{ORDECC-j}{\bm \xi}^{{\bm m}}{\bm \eta}^{{\bm n}}\,,
\end{equation}
where $ORDECC$ indicates the maximal order in eccentricities
considered, here fixed to 12. When the system is far from a
mean-motion resonance, such an analytical approach at first order in the
masses is accurate enough to describe the secular evolution of
extrasolar systems \citep[see  for
  example][]{lib-sans-CeMDA-2013}. The Hamiltonian formulation
Eq. \eqref{eq:sec_ham} has only two degrees of freedom, with the
semi-major axes being constant in the secular approach.

\begin{figure*}[htpb]
  \centering
  \subfloat{\resizebox{0.5\hsize}{!}{\includegraphics[angle=-90]{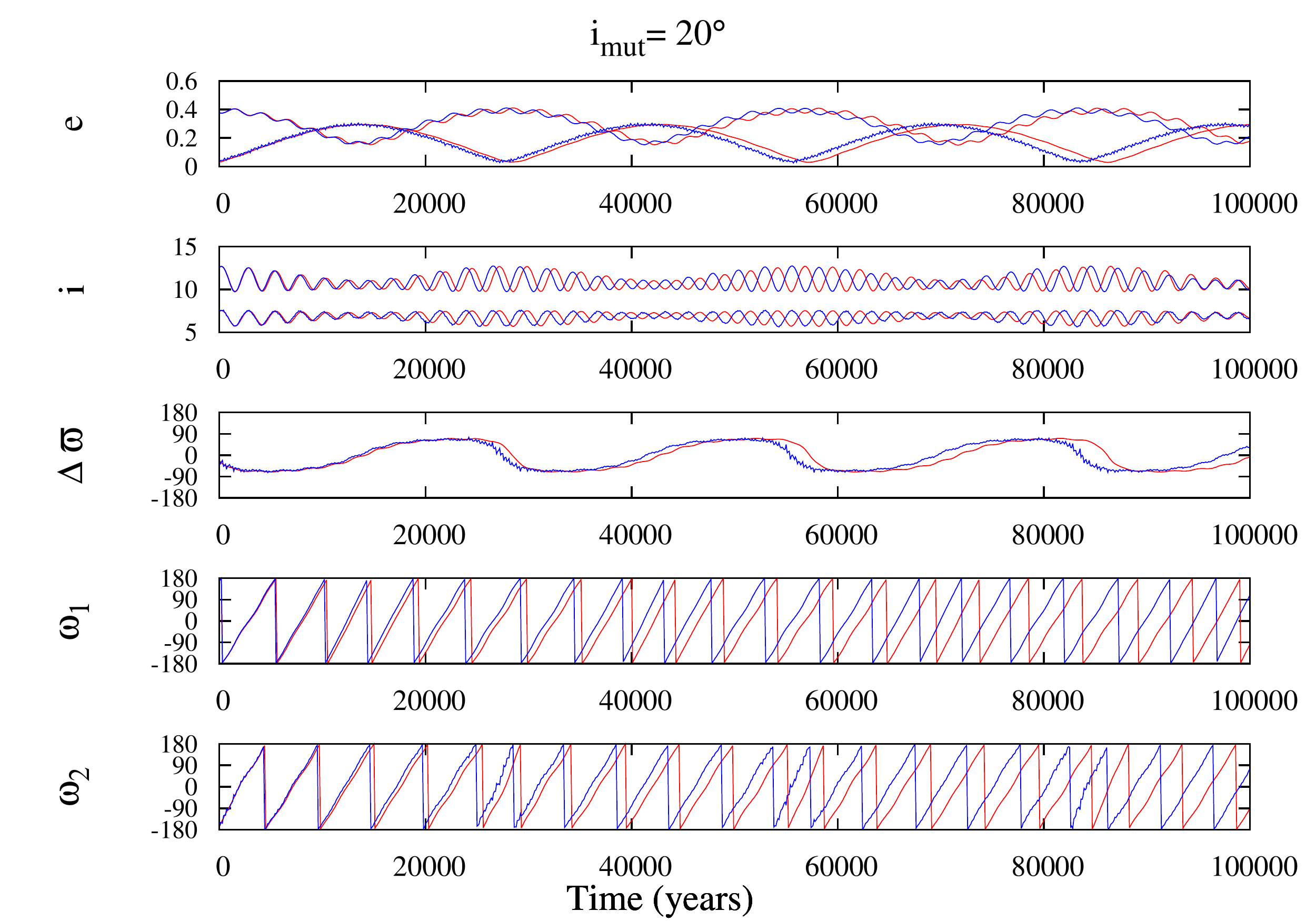}}}
  \subfloat{\resizebox{0.5\hsize}{!}{\includegraphics[angle=-90]{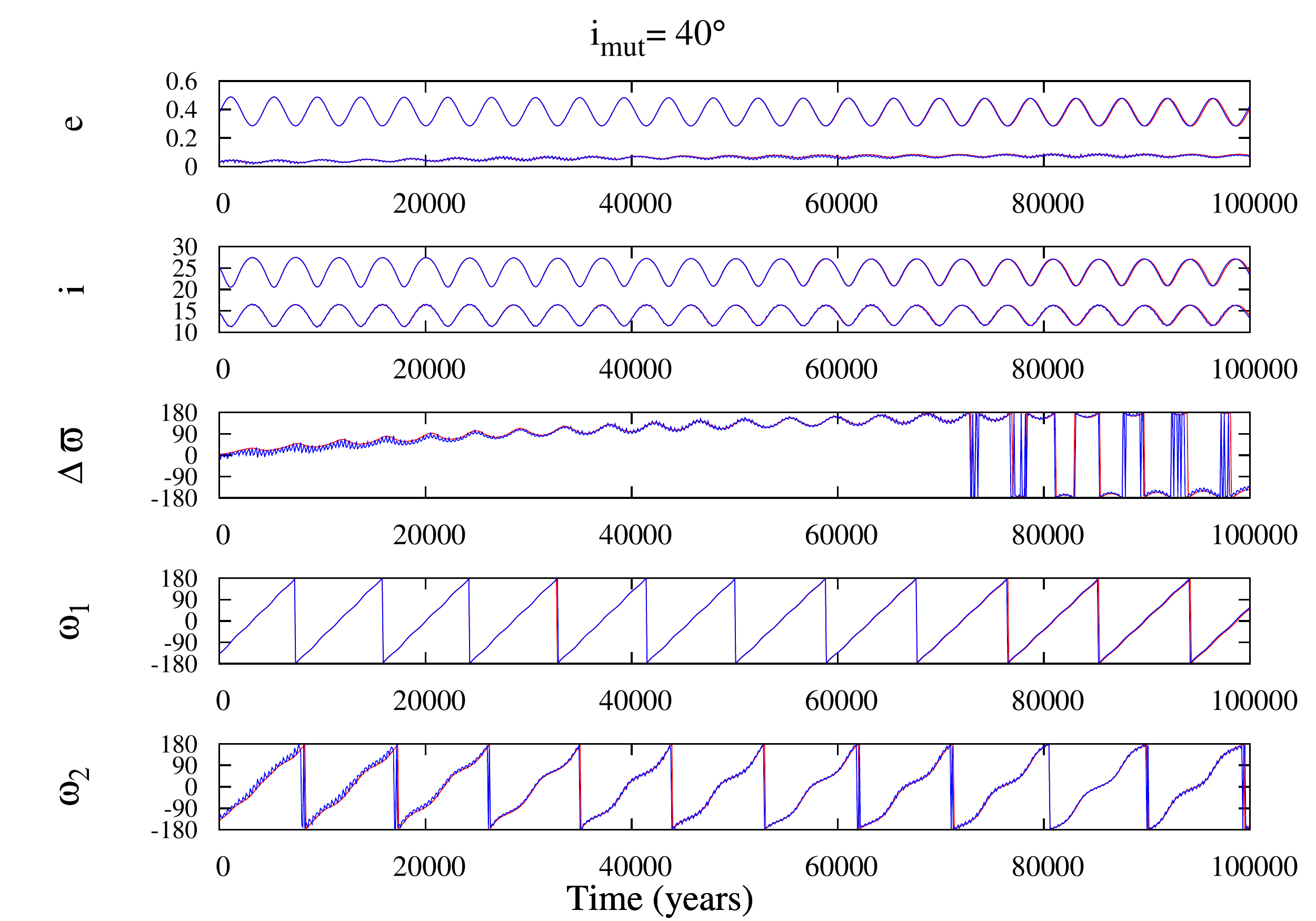}}}\\
  \subfloat{\resizebox{0.5\hsize}{!}{\includegraphics[angle=-90]{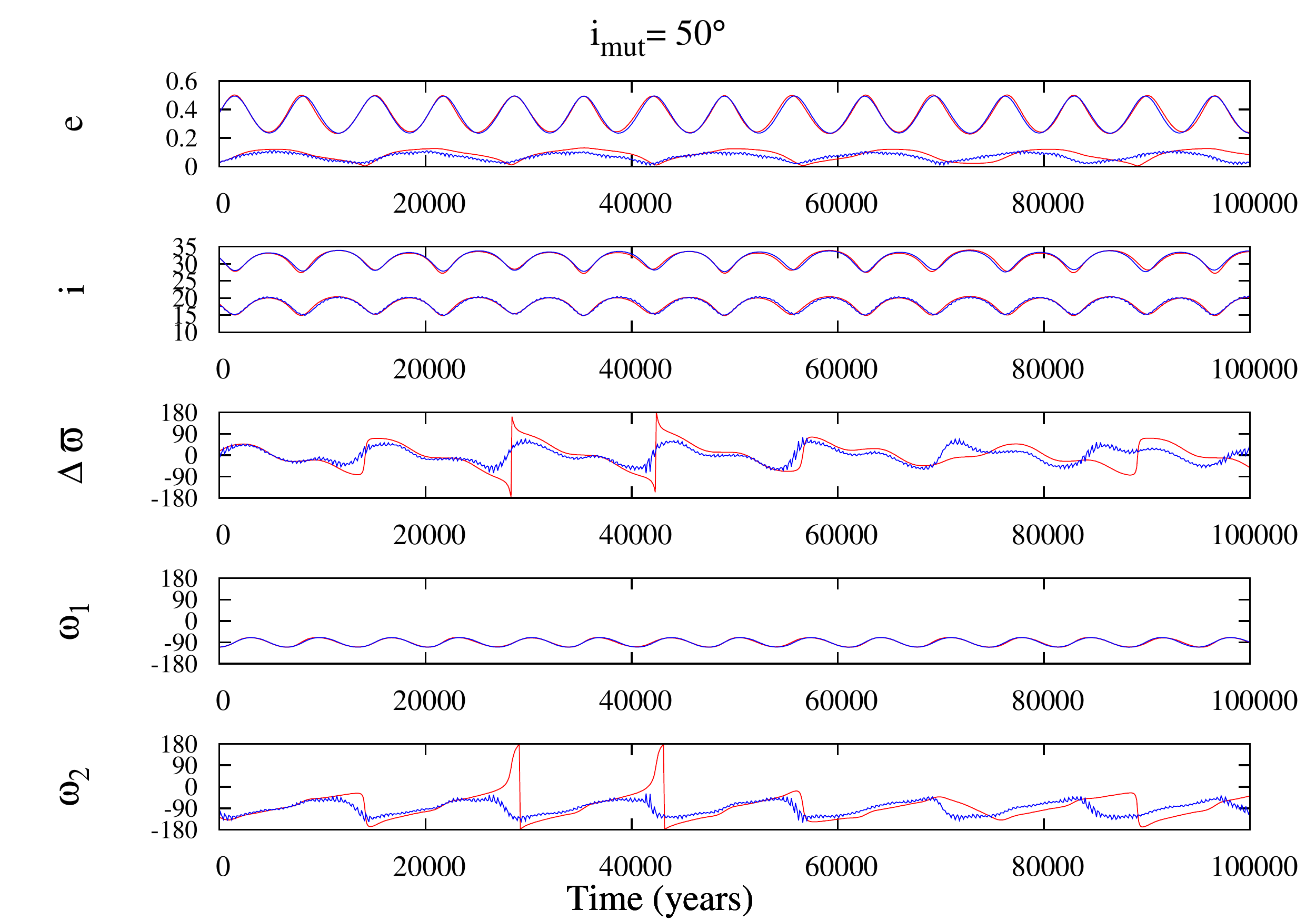}}}
  \subfloat{\resizebox{0.5\hsize}{!}{\includegraphics[angle=-90]{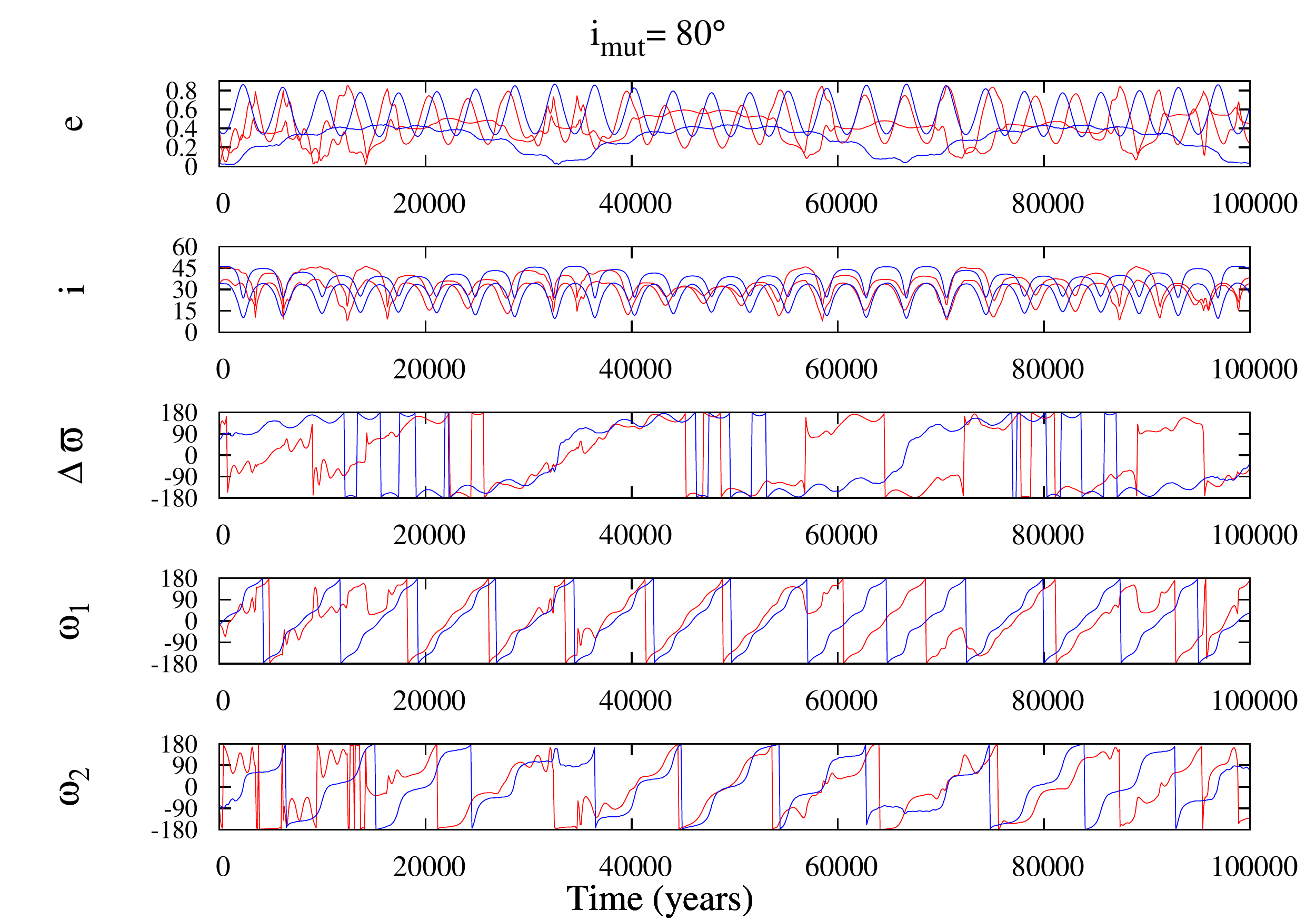}}}
  \caption{Dynamical evolutions of HD~$12661$ system given by the
    analytical expansion (in red) and by n-body simulations (in blue),
    for $i_{mut}=20^\circ$ (top left), $40^\circ$ (top right),
    $50^\circ$ (bottom left), and $80^\circ$ (bottom right). The
    inclination of the orbital plane is fixed to $i = 50^\circ$.}
  \label{fig:HD12661}
\end{figure*}

\section{Parametric study}
\label{sec:parametric}

In the following, we describe the parametric study carried out in the
present work. The selection of the systems considered here is
described in Sect.~\ref{sec:para1}, and the accuracy of the analytical
expansion for the secular evolution of the selected systems is
discussed in Sect.~\ref{sec:para2}.

\subsection{Methodology}\label{sec:para1}
The present work aims to identify the possible 3D architectures of
RV-detected extrasolar systems. From the online database {\tt
  exoplanets.eu}, we selected all the two-planet systems that fulfil
the following criteria: (a) the period of the inner planet is longer
than $45$~days (no tidal effects induced by the star); (b) the
semi-major axis of the outer planet is smaller than $10 \, \text{AU}$
(systems with significant planet-planet interactions); (c) the system
is not close to a mean-motion resonance; (d) the planetary
eccentricities are lower than $0.65$; (e) the masses of the planets
are smaller than $10\,M_J$.
The orbital parameters of the ten selected systems are listed in
Table~\ref{tab:orbpar}, as well as the reference from which they have
been derived.

In this work, the secular evolutions of the systems are considered
when varying the mutual inclination $i_{mut}$ and the orbital plane
inclination $i$ with respect to the plane of the sky. It is important
to note that, although the inclinations $i_1$ and $i_2$ of the two
orbital planes may differ, we decided here to set the same value $i$
for both planes. Thus both masses are varied using the same scaling
factor $\sin i$.

In the general reference frame, the following relation holds:
\begin{equation}\label{eq_imut}
\cos i_{mut} = \cos i_1 \cos i_2 + \sin i_1 \sin i_2 \cos \Delta
\Omega\,,
\end{equation}
being $\Delta \Omega = \Omega_1 - \Omega_2$. It should be noted that
Eq.~\eqref{eq_imut} can be solved if $i_{mut} \leq 2i$, thus for a
given value of $i$ it determines boundaries for the compatible values
of $i_{mut}$. Since $i_1 = i_2 = i$, having fixed the values of
$i_{mut}$, we can determine the value of the longitudes of the nodes
by setting $\Omega_1 = \Delta \Omega$ and $\Omega_2 = 0$, thus
obtaining the complete set of initial conditions. A consequent change
of coordinates to the Laplace plane is finally performed by using the
following relations valid in the Laplace plane:
\begin{equation}
\begin{aligned}
\Lambda_1\sqrt{1-e_1^2}\cos{i_{L1}}+\Lambda_2\sqrt{1-e_2^2}\cos{i_{L2}}=C,\\
\Lambda_1\sqrt{1-e_1^2}\sin{i_{L1}}+\Lambda_2\sqrt{1-e_2^2}\sin{i_{L2}}=0,
\end{aligned}
\end{equation}
where $i_{L1}$ and $i_{L2}$ denote the orbital inclinations in the
Laplace-plane reference frame.

For our parametric study, we varied the value of the mutual
inclination $i_{mut}$ from $0^\circ$ to $80^\circ$ with an increasing
step of $0.5^\circ$, while the common orbital plane inclination $i$
runs from $5^\circ$ to $90^\circ$ with an increasing step of
$5^\circ$. As the coefficients $\mathcal{C}_{j,{\bm m},{\bm n}}$ in
Eq.~\eqref{eq:sec_ham} depend on ${\bm L}$, and therefore on the
masses of the planets, we recomputed them for each value of
$i$. Regarding the integration of the secular approach, we fixed the
integration time to $10^6$ yr with an integration step of~$1$ yr, and
the energy preservation was monitored along the integration.

\subsection{Accuracy of the analytical approach}\label{sec:para2}

Before discussing the results of our parametric study, we need to
ensure that the Hamiltonian formulation,~Eq. \eqref{eq:sec_ham}, provides an
accurate description of the planetary dynamics for all sets of
parameters considered in the study, in particular for high values of
the mutual inclination $i_{mut}$. As already shown in previous papers
(e.g. \cite{lib-hen-cemda-2005} for the coplanar problem,
\cite{lib-hen-icarus-2007} for the 3D problem), the series of the
secular terms converge better than the full perturbation. However, the
higher the value of $D_2$, the weaker the convergence, as expected. In
the following, we discuss the numerical convergence of the expansion
for the selected extrasolar systems, also called {\it convergence au
  sens des astronomes}, as opposed to the mathematical convergence
\citep{poincare-1893}.

\begin{figure*}
\subfloat{\resizebox{\hsize/2}{!}{\includegraphics[angle=-90]{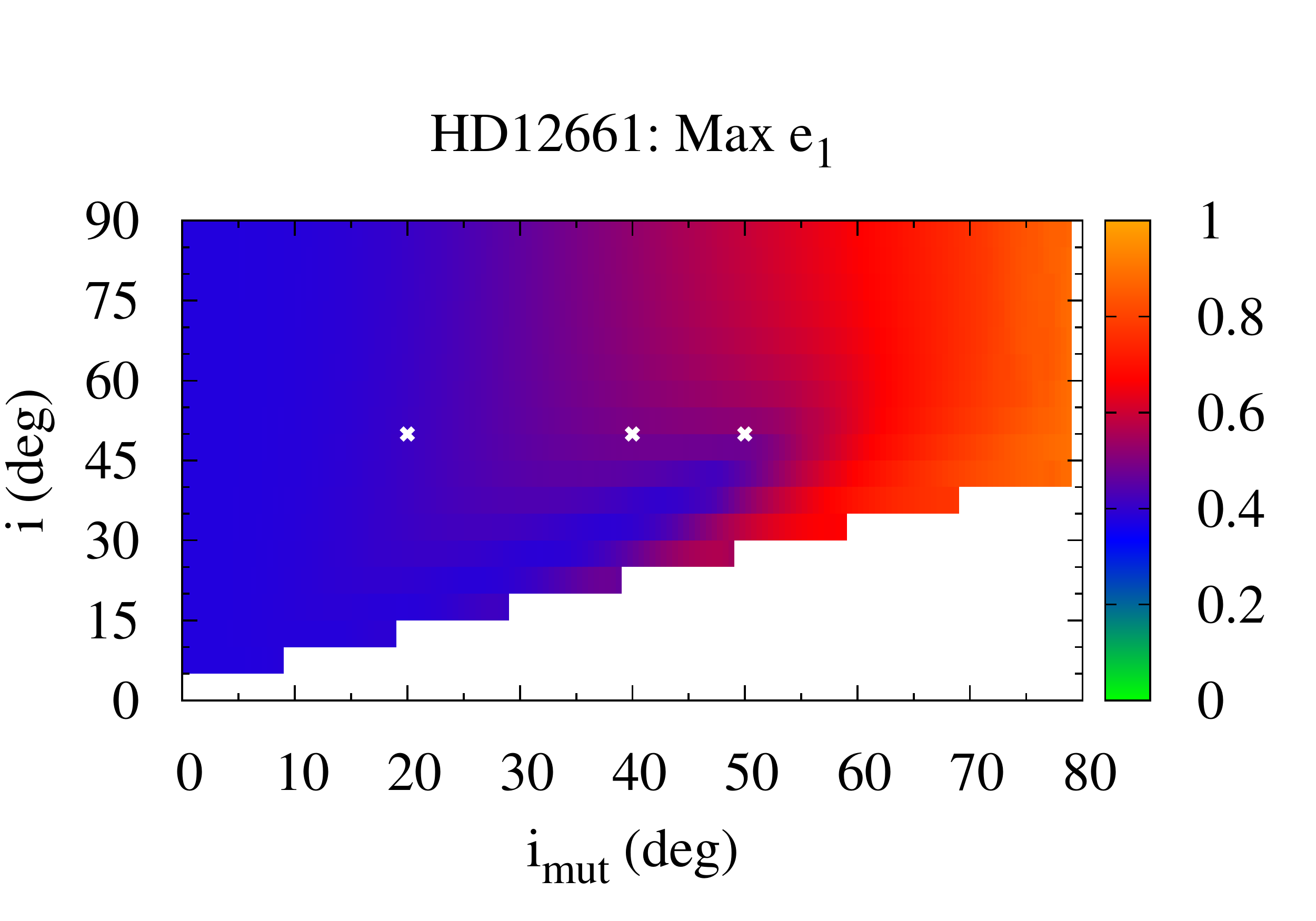}}}
\subfloat{\resizebox{\hsize/2}{!}{\includegraphics[angle=-90]{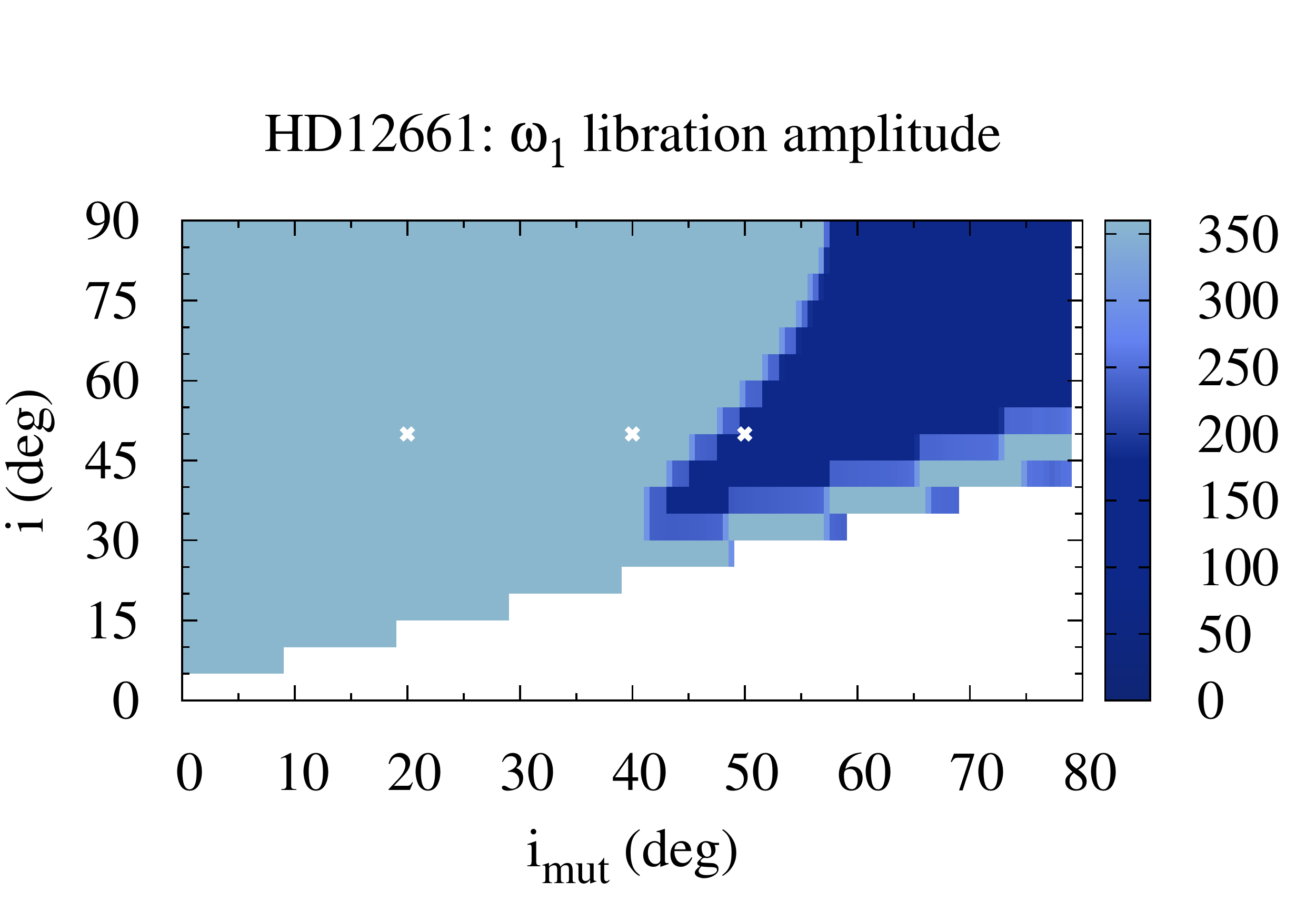}}}
\caption{Long-term evolution of HD~12661 system when varying the
  mutual inclination $i_{mut}$ ($x$-axis) and the inclination of the
  orbital plane~$i$ ($y$-axis), both expressed in degrees. Left panel:
  Maximal eccentricity of the inner planet, as defined by
  Eq.~\eqref{eq:maxe1}. Right panel: Libration amplitude of
  the argument of the pericenter $\omega_1$ (in degrees), as defined
  by Eq.~\eqref{eq:maxomega1}. The three highlighted points are
  related to the representative planes shown in Fig.~\ref{fig:rp}.}
\label{fig:HD12661-LK}
\end{figure*}

\begin{figure*}
  \centering
  \subfloat{\resizebox{\hsize/3}{!}{\includegraphics{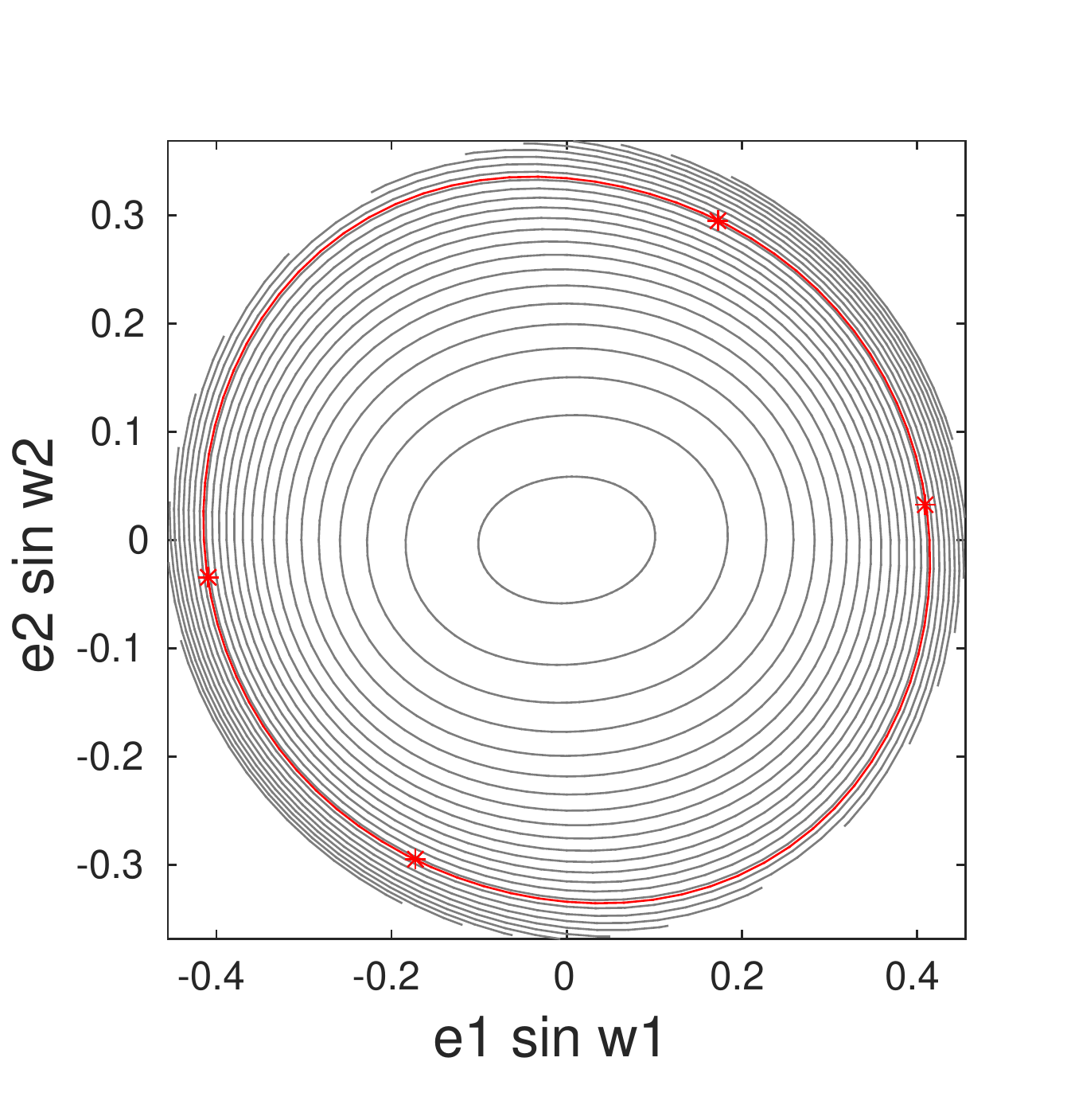}}}
  \subfloat{\resizebox{\hsize/3}{!}{\includegraphics{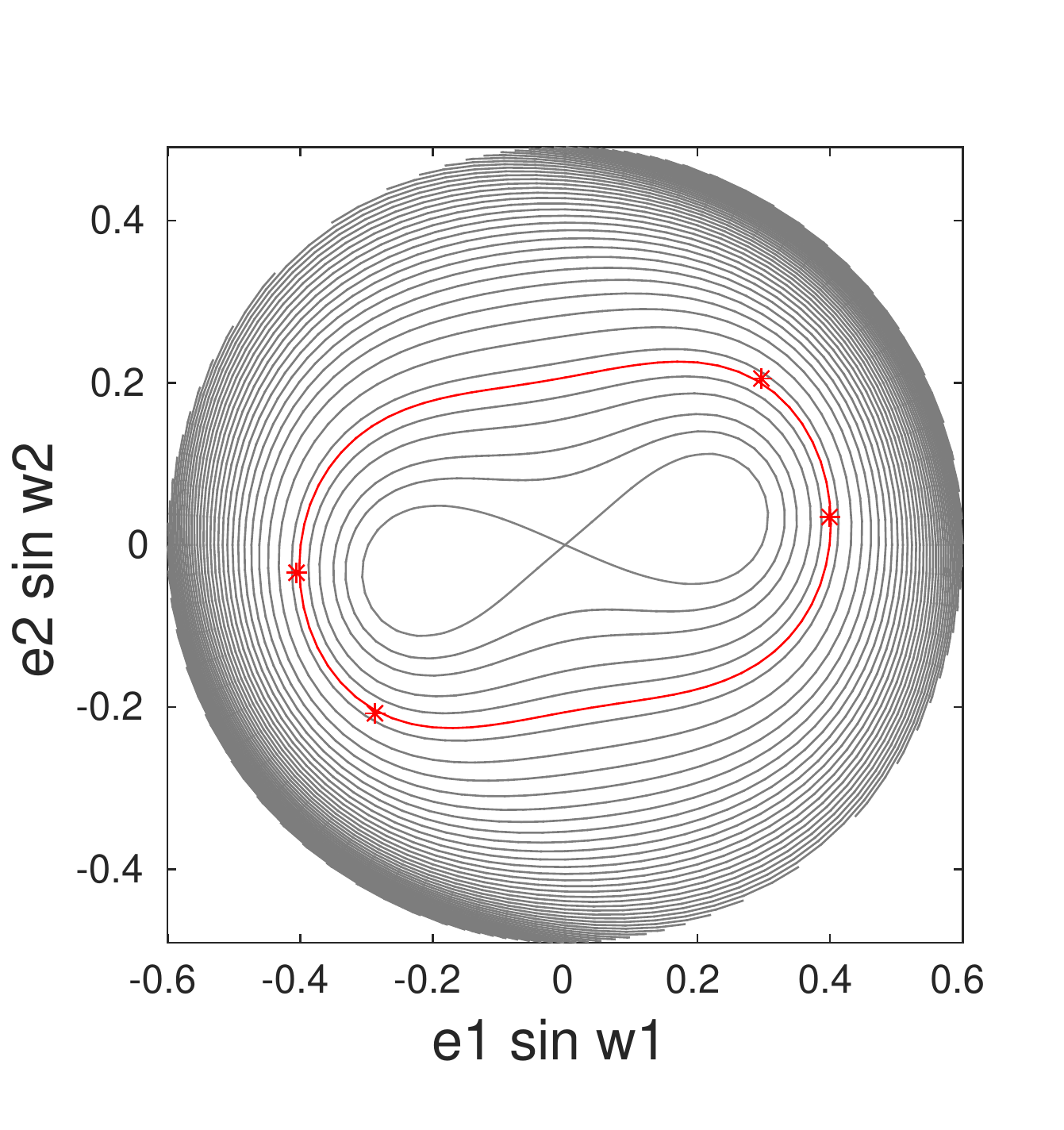}}}
  \subfloat{\resizebox{\hsize/3}{!}{\includegraphics{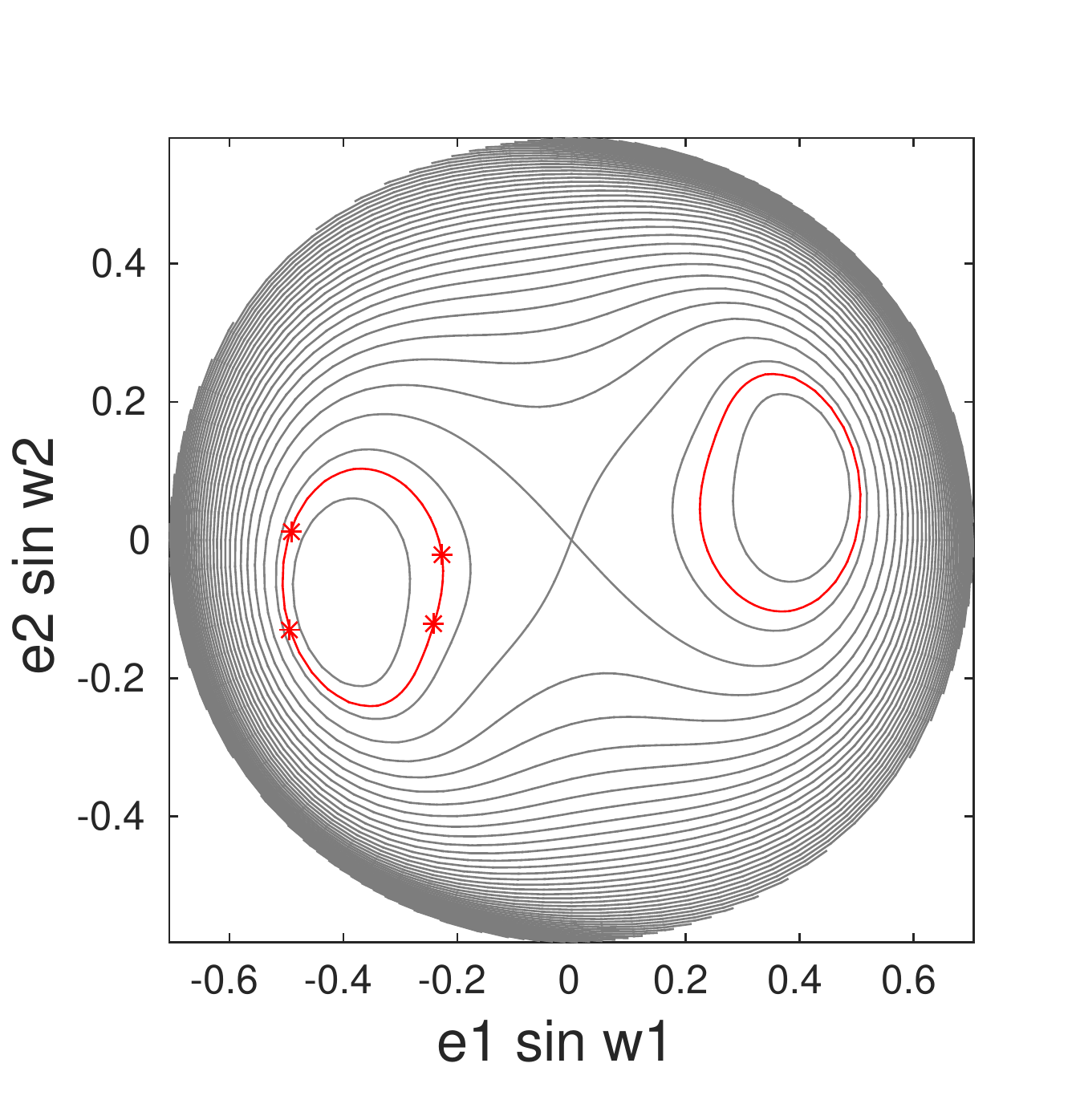}}}
  \caption{Representative plane for HD~12661 system, having fixed the
    inclination of the orbital plane to $i=50^\circ$, for $i_{mut} =
    20^\circ$ (left panel), $i_{mut} = 40^\circ$ (middle panel), and
    $i_{mut} = 50^\circ$ (right panel). The level curve of Hamiltonian
    relative to the orbital parameters of HD~12661 is highlighted in
    red. The crosses indicate the intersections of the orbit with the
    representative plane.}
  \label{fig:rp}
\end{figure*}

Table~\ref{tab:conv} lists, for the ten systems, the contributions to
the Hamiltonian value of the terms from order 2 to order 12 in
eccentricities and inclinations (i.e. $j+m+n$ in
Eq.~(\ref{eq:sec_ham})). The entries are the sums of the terms
appearing at a given order, computed at the orbital parameters given
in Table~\ref{tab:orbpar} and at $i=50^\circ$ and $i_{mut}=50^\circ$,
in order to evaluate the {\it convergence au sens des astronomes} at
high mutual inclination. The numerical convergence of the expansion at
high mutual inclination is obvious for most of the systems. However,
when the decrease of the terms is less marked, we should keep in mind
that results at higher mutual inclinations should be analysed with
caution. Moreover, the last column of Table~\ref{tab:orbpar} gives an
estimation of the remainder of the truncated expansion. It shows the
relative error between the secular Hamiltonian computed by numerical
quadrature and our polynomial formulation (Eq. \ref{eq:sec_ham}),
confirming the previous observations.
 
To further illustrate the accuracy of our analytical approach, we show
in Fig.~\ref{fig:HD12661} the evolutions of HD 12661 given by the
analytical expansion~(Eq. \ref{eq:sec_ham}) (red curves) for the mutual
inclinations $i_{mut}=20^\circ,40^\circ,50^\circ,$ and $80^\circ$ ($i$
is fixed to $50^\circ$), and compare them to the evolutions obtained
by the numerical integration of the three-body problem with the SWIFT
package (\cite{lev-dun-icarus-1994}, blue curves). Although the
numerical convergence observed in Table~\ref{tab:conv} is not
excellent for HD 12661, the agreement of the analytical approach with
the numerical integration of the full problem is very good. The
dynamical evolutions are well reproduced up to high values of the
mutual inclination ($i_{mut}=20^\circ,40^\circ,50^\circ$). Only small
differences in the periods are observed and can be attributed to the
short-period terms not considered in our secular formulation. For very
high values ($i_{mut}=80^\circ$), the dynamical evolutions given by
the two methods no longer coincide, but follow the same trend. As
will be shown in Sect.~\ref{subsec:stability}, the orbits are
generally chaotic at such high mutual inclinations.

\section{Results}
\label{sec:results}

The question of the 3D secular dynamics of RV-detected planetary
systems is addressed here in two directions. Firstly, we focus on
identifying the inclination values for which a LK-resonant regime is
observed in our parametric study. Secondly, the long-term stability of
the mutually inclined systems is unveiled by means of a chaos
detector.

\begin{figure*}
\centering
\subfloat{\resizebox{\hsize/2}{!}{\includegraphics[angle=-90]{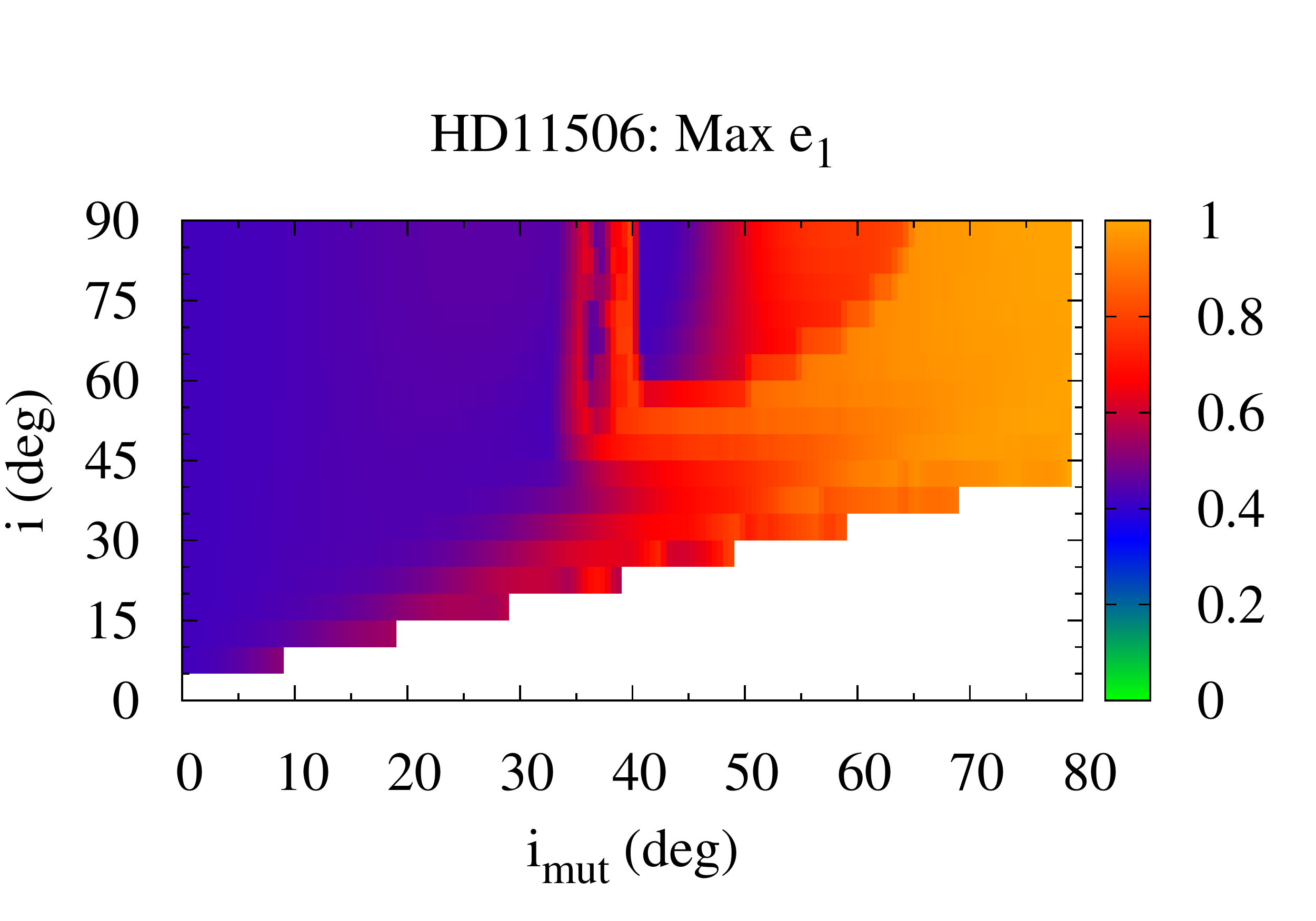}}}
\subfloat{\resizebox{\hsize/2}{!}{\includegraphics[angle=-90]{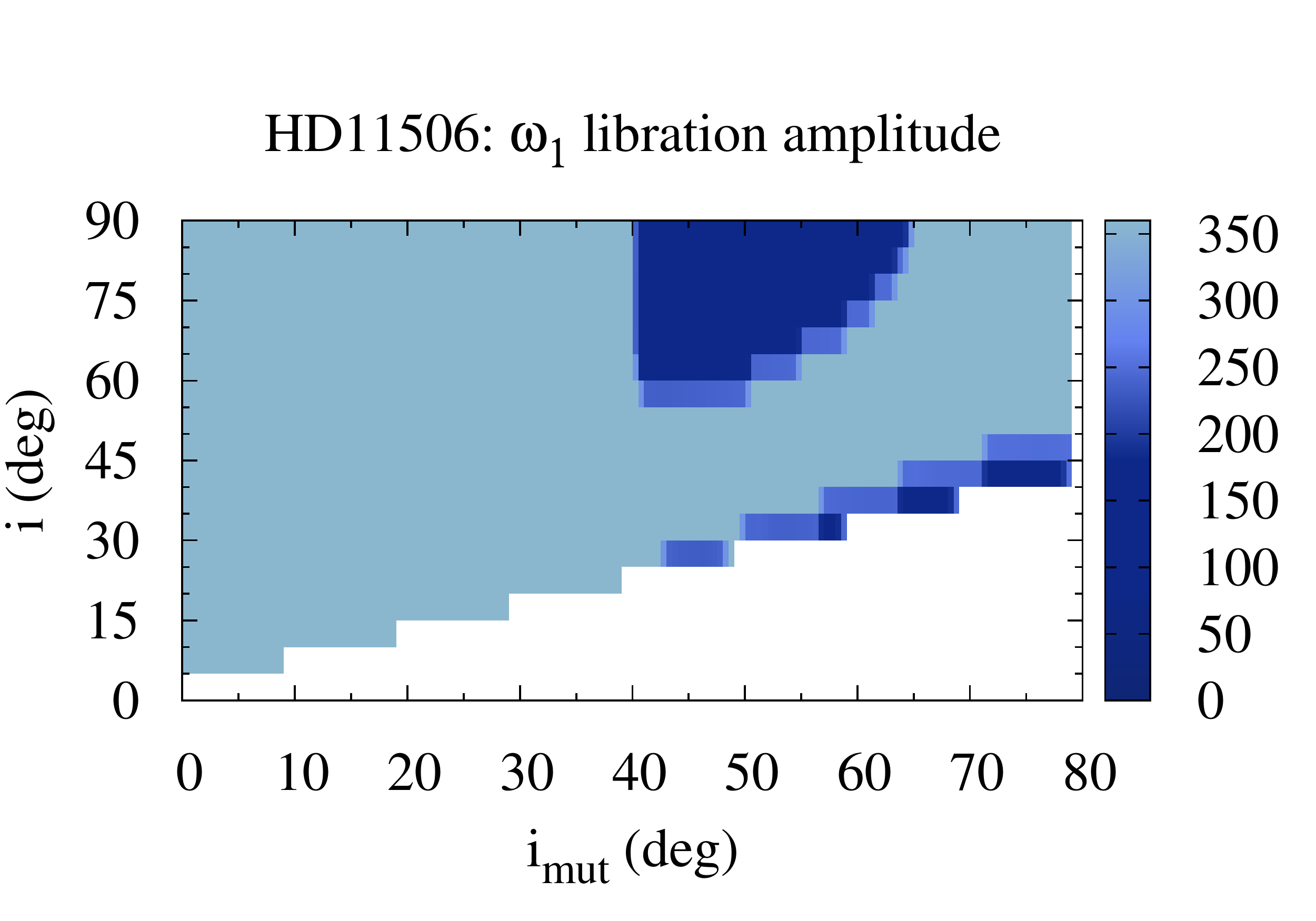}}}
\caption{Same as Fig.~\ref{fig:HD12661-LK} for HD~$11506$ system.}
\label{fig:HD11506-LK}
\end{figure*}

\subsection{Extent of the Lidov-Kozai regions}
\label{subsec:reskozai}

Regarding the possible 3D configurations of extrasolar systems, we are
particularly interested in the LK resonance. This protective mechanism
ensures that the system remains stable, despite large eccentricities
and inclination variations. It is characterised, in the Laplace-plane
reference frame, by the coupled variation of the eccentricity and the
inclination of the inner planet, and the libration of the argument of
the pericenter of the same planet around $\pm90^{\circ}$
\citep{Lidov1962,kozai-1962}.

As a first example, we investigate the dynamics of the HD~12661 extrasolar
system. In the left panel of Fig.~\ref{fig:HD12661-LK}, we show, for
varying ($i_{mut},i$) values, the maximal eccentricity of the inner
planet reached during the dynamical evolution of the system,
\begin{equation}
  \label{eq:maxe1}
\max e_1 = \max_t e_1(t)\,,
\end{equation}
being $e_1(t)$ the eccentricity of the inner planet at time $t$.  Let
us note that this quantity is often used to determine the regularity
of planetary orbits, since for low ($e<0.2$) and high ($e>0.8$)
eccentricity values it is generally found to be in good agreement with chaos
indicators \citep[see for instance][]{funk-et-al-AA-2011}. On the
right panel, we report, for all the considered ($i_{mut},i$) values,
the libration amplitude of the angle $\omega_1$, defined as
\begin{equation}
  \label{eq:maxomega1}
{\rm libr\_ampl}\,(\omega_1) = \max_t\omega_1(t) -\min_t\omega_1(t)\,.
\end{equation}
This value will serve as a guide for the detection of the LK-resonant
behaviour characterised by the libration of $\omega_1$, and thus by a
small value of ${\rm libr\_ampl}\,(\omega_1)$.

When following an horizontal line in Fig.~\ref{fig:HD12661-LK}, the
mutual inclination $i_{mut}$ varies while the orbital inclination $i$,
and thus the planetary masses, are kept fixed. On the other hand, the
inclination of the common orbital plane decreases when moving down
along a vertical line, while the planetary masses increase
accordingly. As previously stated, this implies the recomputation of
the coefficients $\mathcal{C}_{j,{\bm m},{\bm n}}$ of
Eq.~\eqref{eq:sec_ham}. Let us recall that all ($i_{mut}$, $i$) pairs
cannot be considered here since, for fixed $i_1=i_2=i$ values,
Eq.~(\ref{eq_imut}) cannot be solved for all the mutual inclinations.

We see that the eccentricity variations of 3D configurations of HD
12661 are small\footnote{The initial inner
  eccentricity is $0.377$.} for low mutual inclinations (blue
in the left panel of Fig.~\ref{fig:HD12661-LK}) and become large for
high mutual inclinations (red). Additionally, the argument of
the pericenter $\omega_1$ circulates for low $i_{mut}$ values (light
blue in the right panel of Fig.~\ref{fig:HD12661-LK}) and
librates for high $i_{mut}$ values (dark blue). Thus, for high
mutual inclinations, the system is in a LK-resonant state.

To visualise the different dynamics, we draw, for a given $D_2$ value,
the level curves of the Hamiltonian~(Eq. \ref{eq:sec_ham}) in the
representative plane $(e_1 \sin{\omega_1},e_2 \sin{\omega_2})$ where
both pericenter arguments are fixed to $\pm 90^{\circ}$ (see
\cite{lib-hen-icarus-2007} for more details on the representative
plane). This plane is neither a phase portrait nor a surface of
section, since the problem is four dimensional. However, nearly all
the orbits will cross the representative plane at several points of
intersection on the same energy curve. Figure~\ref{fig:rp} shows the
representative planes of HD~12661 for $i_{mut}=20^{\circ}$ (i.e.
$D_2=0.35$, left panel), $40^{\circ}$ (i.e. $D_2=0.67$, middle panel),
and $50^{\circ}$ (i.e. $D_2=0.90$, right panel), the inclination of
the orbital plane being fixed to $50^\circ$. These three system
configurations are also indicated with white crosses in
Fig.~\ref{fig:HD12661-LK} and their dynamical evolutions are those
presented in Fig.~\ref{fig:HD12661}.

For low values of $i_{mut}$, circular orbits ($e_1 = e_2 = 0$)
constitute a point of stable equilibrium (left panel of
Fig.~\ref{fig:rp}). As we increase the mutual inclination (central and
right panels of Fig.~\ref{fig:rp}), the central equilibrium becomes
unstable and bifurcates into the two stable LK equilibria. The red
crosses represent the intersections of the evolution of the mutually
inclined HD 12661 system with the representative plane. For low mutual
inclinations, the crosses are located on both sides of the
representative plane, so the argument of the inner pericenter
circulates. For $i_{mut}=50^{\circ}$ (right panel of
Fig.~\ref{fig:rp}), the crosses are inside the LK island in the left
side of the representative plane, associated with the libration of
$\omega_1$ around $270^{\circ}$ (as can also be observed in the
bottom left dynamical evolution shown in Fig.~\ref{fig:HD12661}). We
see that the corresponding white cross on the right side of
Fig.~\ref{fig:HD12661-LK} is likewise located inside the dark blue
region of the LK resonance.

The critical value of the mutual inclination, which corresponds to the
change of stability of the central equilibrium, depends on the mass
and semi-major axis ratios \citep[see \eg][]{lib-hen-icarus-2007} and
is typically around $40^{\circ}-45^{\circ}$ for mass ratios between
0.5 and 2. For increasing mutual inclinations, the stable LK
equilibria reach higher inner eccentricity values and the orbit of the
considered system possibly crosses the representative plane inside a
LK island. Therefore, the dark blue LK region in
Fig.~\ref{fig:HD12661-LK} starts around $40^{\circ}-55^\circ$, the
exact value for the change of dynamics depending on the inclination of
the orbital plane since the expansion~(Eq. \ref{eq:sec_ham}) depends on
the inclination $i$ via the planetary mass.

Let us note that, even if the numerical convergence of the analytical
expansion of the HD~12661 system is not excellent (see
Table~\ref{tab:conv}), the LK-resonant region perfectly matches the
one obtained with n-body simulations additionally performed for
validation, except at very high mutual inclinations
($i_{mut}\ge70^\circ$). Indeed, for the HD~12661 and HD~74156 systems, a
destabilisation of the orbits is observed at very high mutual
inclinations and slightly reduces the stable LK region.

A second example is shown in Fig.~\ref{fig:HD11506-LK} for the HD~11506
system. The LK region is now located at smaller mutual inclinations,
making visible the right border of the LK region. For each $i$ value,
the interval of mutual inclinations associated with the libration of
the angle $\omega_1$ begins at $\sim 40^\circ$, whereas its amplitude
depends on $i$. No spatial configuration of HD~11506 can be found in a
LK-resonant state for a mutual inclination higher than $65^\circ$.

Let us note that some additional dark blue points can be observed
for low values of the inclinations of the orbital plane $i$. These
systems are close to the separatrix of the LK resonance and will be
destabilised on a longer timescale, as will be shown in the next
section.

\begin{table}
\centering
\caption{Extent of the LK region for the ten systems. For each system,
  we indicate the minimum $i_{mut}$ (second column) and $i$ (third
  column) values of the LK region where libration of $\omega_1$ is
  observed in Fig.~\ref{fig:lidov-kozai}, the percentage of initial
  conditions for which a LK-resonant state is observed (fourth column),
  and the percentage of initial conditions classified as chaotic by
  the chaos indicator (fifth column).}
\label{tab:res}
\begin{tabular}{lcccccc}
\hline\hline System & $\min i_{mut}$ & $\min i$ & LK & chaos \\ &
$(^\circ)$ & $(^\circ)$ & (\%) & (\%) \\
\hline
HD 11506 & 41 & 30 & 15 & 39 \\
HD 12661 & 43 & 30 & 24 & 49 \\
HD 134987 & 46 & 30 & 13 & -- \\
HD 142 & 44 & 30 & 11 & 2 \\
HD 154857 & 41 & 30 & 10 & 2 \\
HD 164922 & 43 & 30 & 23 & -- \\
HD 169830 & 45 & 25 & 23 & 19 \\
HD 207832 & 50 & 35 & 17 & 20 \\
HD 4732 & 49 & 35 & 12 & 15 \\
HD 74156 & 41 & 30 & 20 & 2 \\
\hline
\end{tabular}
\end{table}

\begin{figure*}
\centering \rotatebox{270}{ \subfloat{\resizebox{0.7\hsize}{!}
    {\includegraphics{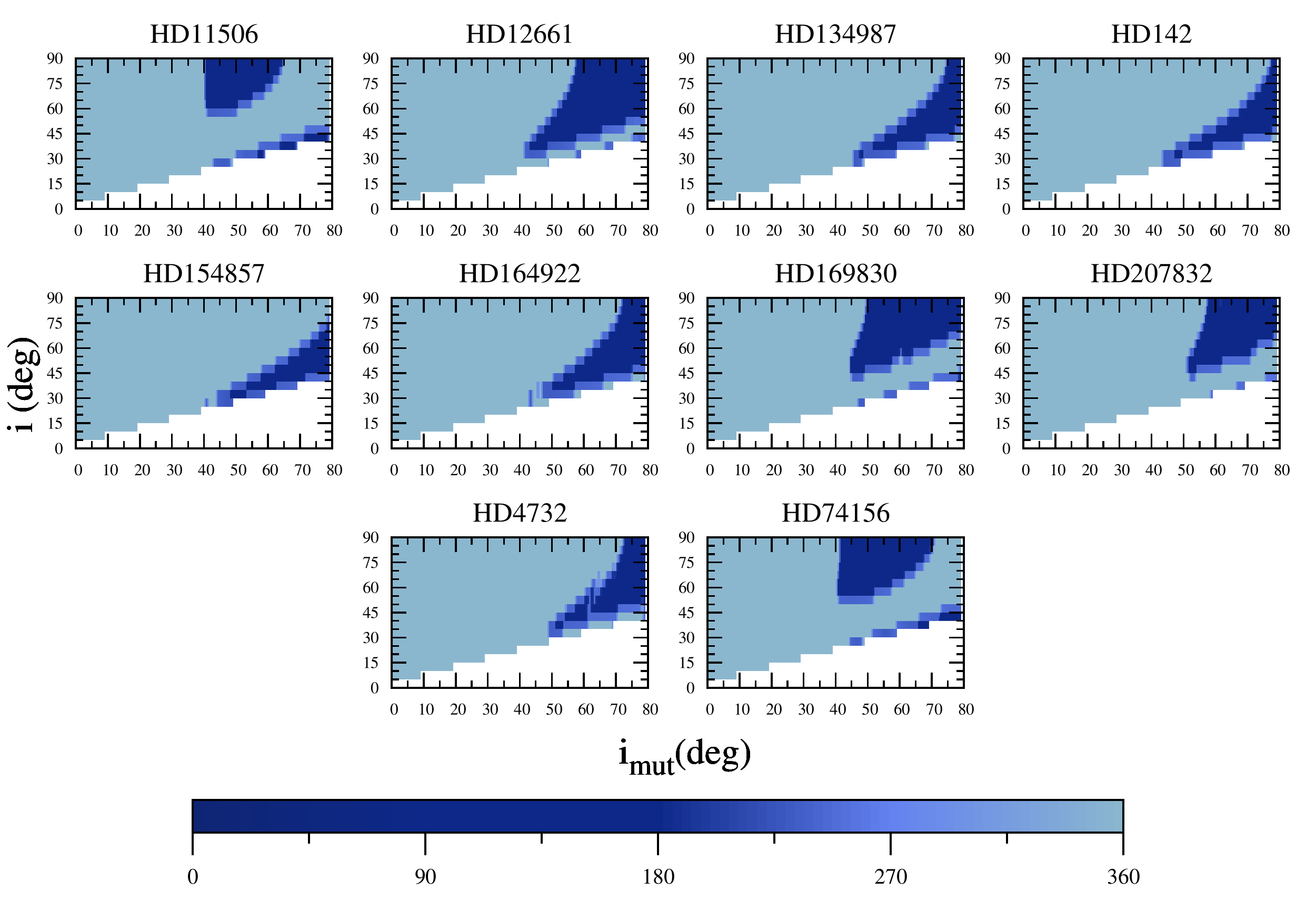}}}}
\caption{Libration amplitude of $\omega_1$ for the ten systems
  considered here, when varying the mutual inclination $i_{mut}$
  ($x$-axis) and the inclination of the orbital plane~$i$ ($y$-axis).}
 \label{fig:lidov-kozai}
\end{figure*}

In Fig.~\ref{fig:lidov-kozai} we display the libration amplitude
of the argument of the pericenter of the inner planet for the ten
systems considered here. All the graphs do show a LK region. In
other words, all the selected RV-detected systems, when
considered with a significant mutual inclination, have physical and
orbital parameters compatible with a LK-resonant
state. Table~\ref{tab:res} summarises information on the extent of the
LK region for each system. The second and third columns display the
minimum values of the mutual inclination $i_{mut}$ (with an accuracy
of $1^\circ$) and the orbital inclination $i$, respectively, for which
a libration of the argument of the pericenter $\omega_1$ is
observed. The percentage of initial conditions inside the (dark blue)
LK region is given in the fourth column. The last column reports the
percentage of chaos in the whole set of initial conditions and will be
discussed in Sect.~\ref{subsec:stability}.

\begin{figure*}
\centering
\includegraphics[width=12cm,angle=-90]{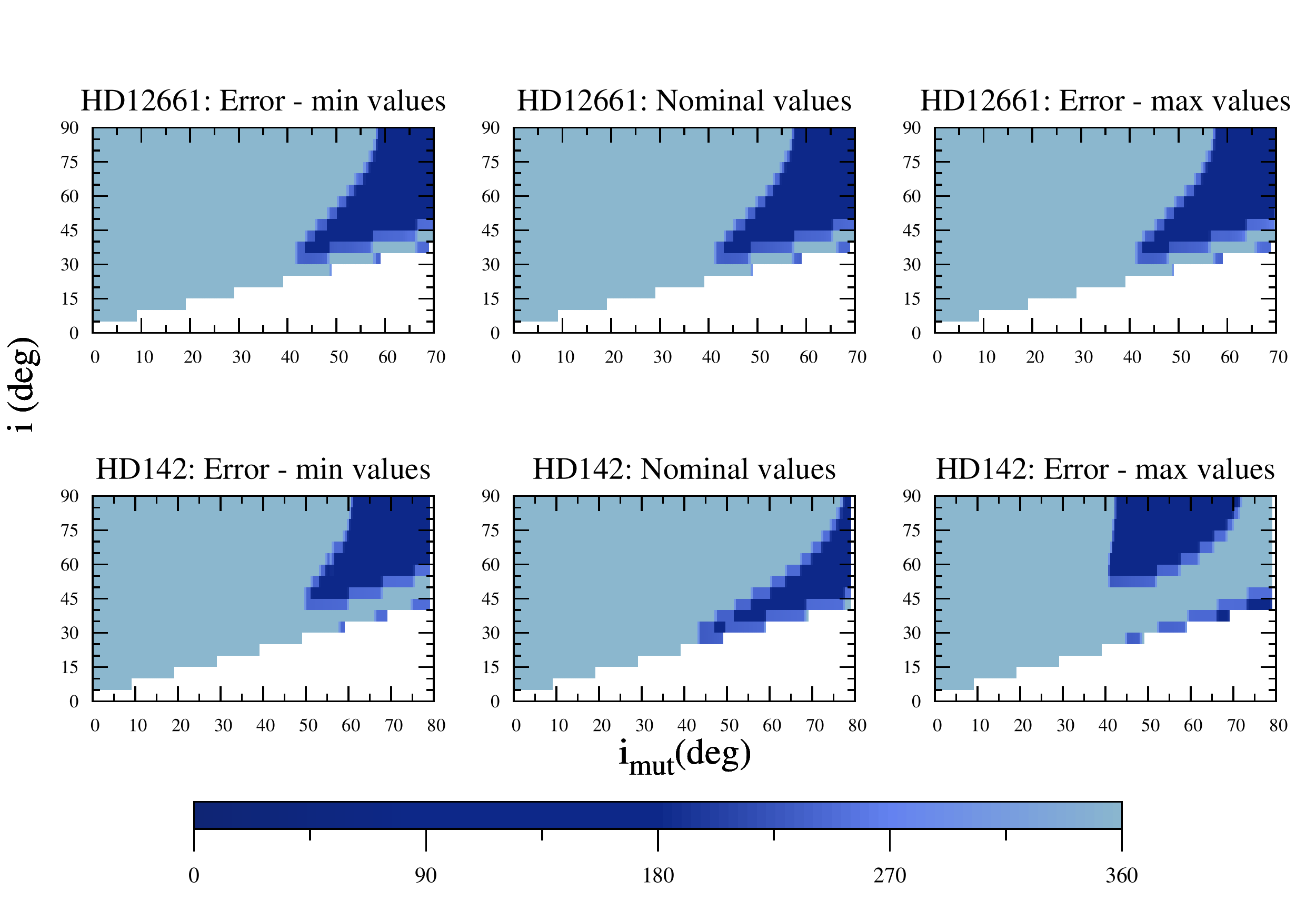}
\caption{Libration amplitude of $\omega_1$, as in
  Fig.~\ref{fig:lidov-kozai}, for the HD~$12661$ (top) and HD~$142$
  (bottom) systems, when considering the minimal values (left), the
  nominal values (middle), and the maximal values (right) of the
  orbital parameters.}
\label{fig:max_libr_w1_errors}
\end{figure*}

\subsection{Sensitivity to observational uncertainties}
\label{subsec:sensitivity}
So far, we have considered the nominal values of the orbital
parameters given by the observations. However, due to the limitations
of the detection techniques, observational data come with relevant
uncertainties, and to explore the influence of such uncertainties on
the previous results is relevant. As typical examples, we show in
Fig.~\ref{fig:max_libr_w1_errors} the extent of the LK region for
the HD~$12661$ and HD~$142$ systems, when considering extremal orbital
parameters within the confidence regions given by the observations,
instead of the best-fit parameter values. The errors on each orbital
parameter are listed in Table~\ref{tab:orbpar} for both planetary
systems. Two extremal cases are examined in the following, where the
minimal/maximal values are adopted for all the parameters
simultaneously.

In the case of the HD~$12661$ system, the location and extent of the LK
region are very similar when adopting the minimal values (top left
panel of Fig.~\ref{fig:max_libr_w1_errors}), the nominal values (top
middle), and the maximal values (top right) of the orbital
parameters. Concerning HD~$142$, the situation is quite different. We
observe, in the bottom panels of Fig.~\ref{fig:max_libr_w1_errors}, a
significant variation of the LK region in its extent and shape,
probably due to the greater size of the observational errors on the
different orbital elements. 

As a result, the location and extent of the LK resonance regions are
sensitive to observational uncertainties in the orbital elements,
especially when they are significant, and this should be taken into
account in detailed studies of the selected systems. Nevertheless, we
stress that, when considering extremal values within the confidence
regions, the dynamics remains qualitatively the same, with the
existence of stable LK islands at high mutual inclinations for both
systems.

\begin{figure*}
\centering
\subfloat{\resizebox{\hsize/2}{!}{\includegraphics[angle=-90]{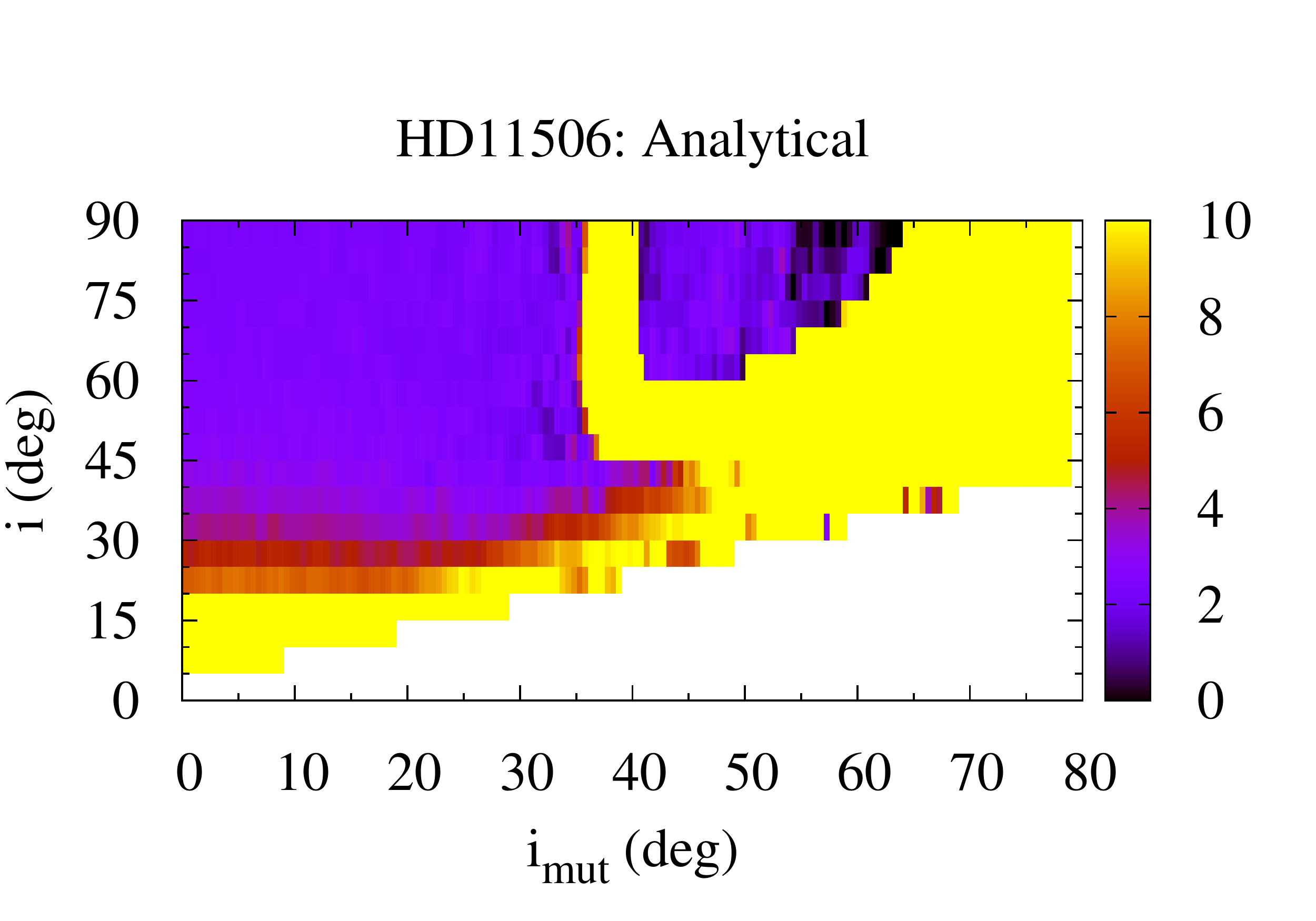}}}
\subfloat{\resizebox{\hsize/2}{!}{\includegraphics[angle=-90]{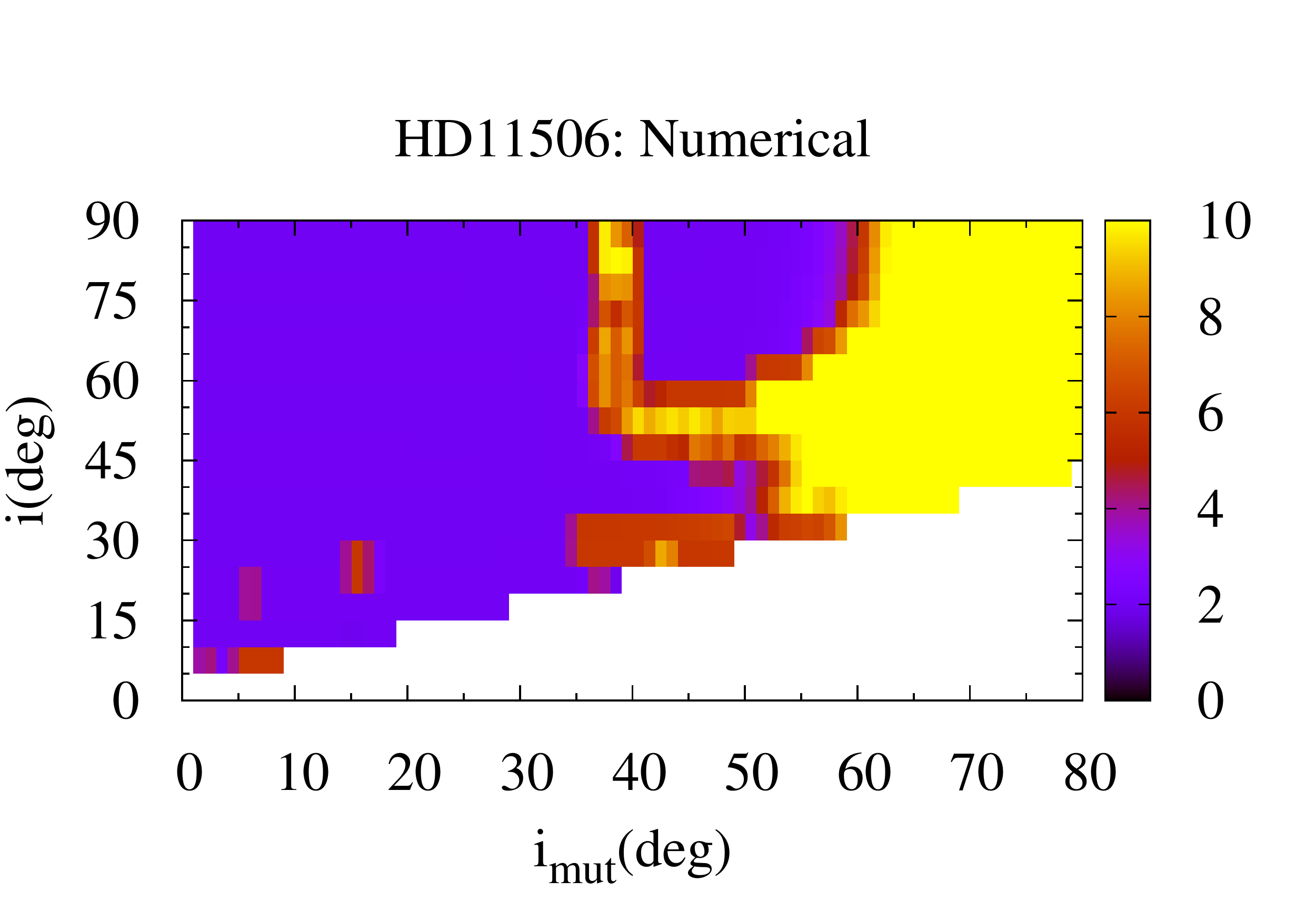}}}
\caption{Mean MEGNO values for the HD~$11506$ system given by our
  analytical approach (left panel) and n-body simulations (right
  panel).}
\label{fig:HD11506-megno}
\end{figure*}

\begin{figure*}
\centering \rotatebox{270}{ \subfloat{\resizebox{0.7\hsize}{!}
    {\includegraphics{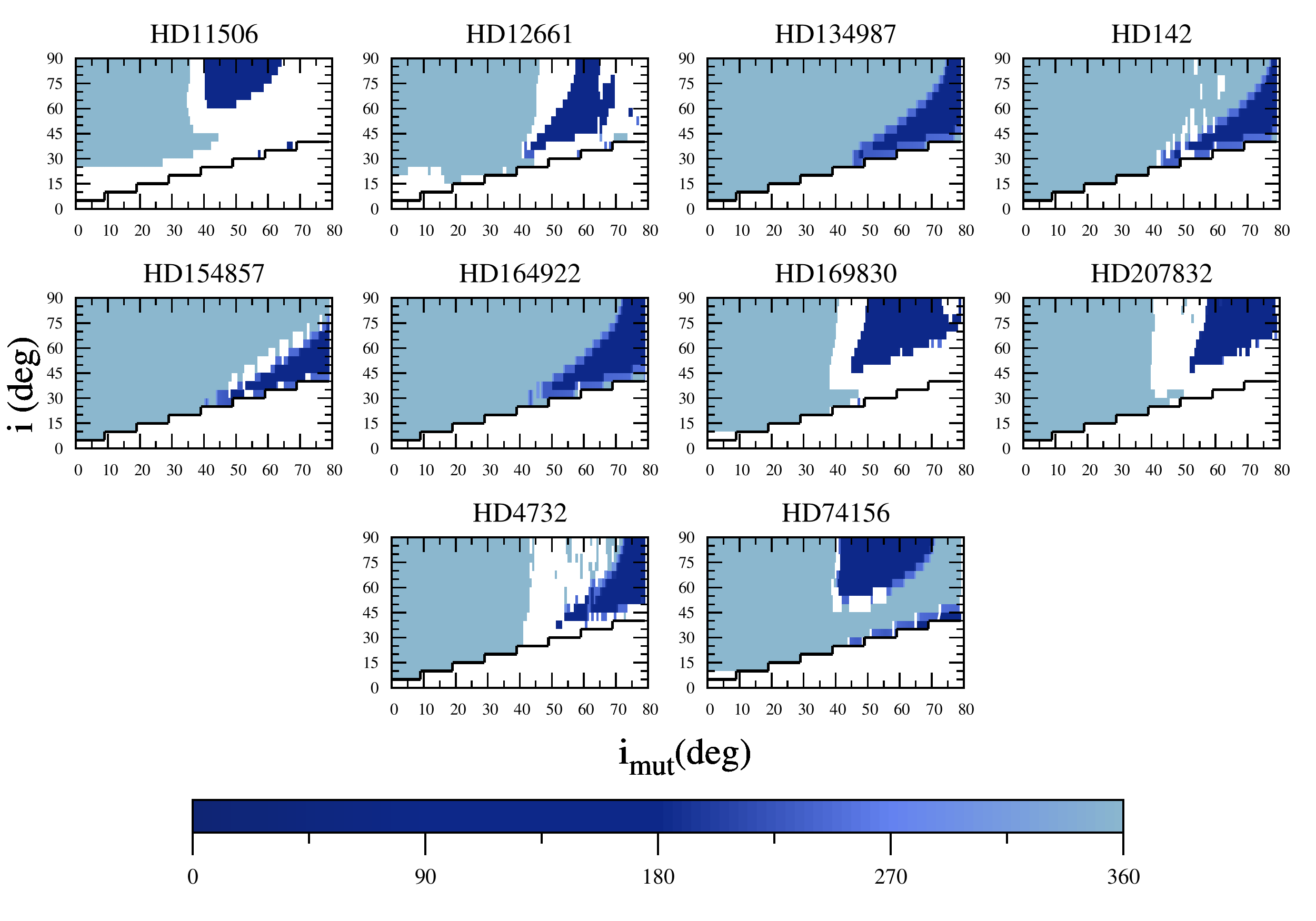}}}}
\caption{Same as Fig.~\ref{fig:lidov-kozai}, where the initial system
  parameters leading to chaotic motion (defined by the mean MEGNO
  value greater than $8$) are coloured in white (above the black
  curve).}
 \label{fig:LK_megno}
\end{figure*}

\subsection{Stability of planetary systems}
\label{subsec:stability}

In this section, we aim to determine if the LK-resonant state of a 3D
planetary system is essential to ensure its long-term stability. To do
so, we have used the Mean Exponential Growth factor of Nearby Orbits
(MEGNO) chaos indicator, briefly described in the following \citep[for
  an extensive discussion on the properties of the MEGNO, see
][]{cin-simo-AASS-2000,maf-gio-cin-2011}.

Let $H(\mathbf{p},\mathbf{q})$ with $\mathbf{p},\mathbf{q} \in
\mathbb{R}^N$ be an autonomous Hamiltonian of $N$ degrees of
freedom. The Hamiltonian vector field can be expressed as
\begin{equation}
\dot{\mathbf{x}} = J\nabla_{\mathbf{x}}\mathcal{H}\mathbf{x}
,\end{equation}
where $\mathbf{x} = \left(%
\begin{array}{c}
\mathbf{p}\\ \mathbf{q}
\end{array}
\right) \in \mathbb{R}^{2N}$ and
J = $\begin{bmatrix} 0_N & -1_N\\ 1_N & \phantom{-}0_N
\end{bmatrix}\,$,
being $1_N$ and $0_N$ the unitary and null $N \times N$ matrices,
respectively.  In order to apply the MEGNO chaos indicator, we need to
compute the evolution of deviation vectors $\bm \delta(t)$. These
vectors satisfy the variational equations
\begin{equation}
\dot{\bm\delta}(t) =
J\nabla_{\mathbf{x}}^2\mathcal{H}\bm\delta(t),
\end{equation}
being $\nabla_{\mathbf{x}}^2\mathcal{H}$ the Hessian matrix of the
Hamiltonian. As in~\citet{cin-simo-AASS-2000}, the Mean Exponential
Growth Factor is defined as
\begin{equation}
Y(t) = \frac{2}{t} \int_{0}^{t} \frac{\dot{\delta}(s)}{\delta(s)}\, ds
,\end{equation}
where $\delta(s)$ is the Euclidean norm of $\bm \delta(s)$. We
consider here the mean MEGNO, that is, the time-averaged MEGNO,
\begin{equation}
\label{eq:megno}
\bar{Y}(t) = \frac{1}{t} \int_{0}^{t} Y(s)\, ds .
\end{equation}
The limit for $t \rightarrow \infty$ provides a good characterisation
of the orbits. The MEGNO chaos indicator is particularly convenient
since we have:
\begin{itemize}
\item $lim_{t \rightarrow \infty} \bar{Y}(t) = 0$ for stable periodic
  orbits,
\item $lim_{t \rightarrow \infty} \bar{Y}(t) = 2$ for quasi-periodic
  orbits and for orbits close to stable periodic ones,
\item for irregular orbits, $\bar{Y}(t)$ diverges with time.
\end{itemize}
For each set of initial conditions we choose the initial deviation
vector $\bm\delta(0)$ as a random unitary vector. We then study
its evolution along the orbit and compute the corresponding evolution
of the mean MEGNO. Two main factors have motivated the choice of this
chaos indicator. First, it requires the study of the evolution of only
one deviation vector, saving valuable computational time. Second, it
returns an absolute value, as it classifies each orbit independently.

As previously noted, the LK-resonant state is surrounded by a chaotic
zone associated with the bifurcation of the central equilibrium at
null eccentricities. Therefore, a chaos indicator can be useful to
highlight the extent of the chaotic zone and identify with precision
the ($i_{mut}$, $i$) values ensuring the regularity of the orbits for
a long time.

On the left panel of Fig.~\ref{fig:HD11506-megno}, we show the values
of the mean MEGNO for HD~$11506$ computed with our analytical
approach. We can appreciate how the region at high inclinations
characterised as regular by the mean MEGNO (purple) clearly
superimposes with the LK-resonant region identified in
Fig~\ref{fig:HD11506-LK}. The surrounding chaotic region displayed in
yellow extends up to high mutual inclinations, showing that highly
mutually inclined configurations of the HD~11506 system can only be
expected in a LK-resonant state. Regarding low mutual inclinations,
nearly all spatial configurations present regular motion up to a
mutual inclination of $\sim35^{\circ}$, where the LK resonance comes
into play.

A comparison with n-body simulations (short-period effects included)
is given in the right panel of Fig.~\ref{fig:HD11506-megno}, where
numerical integrations have been carried out with SWIFT (for every
$1^\circ$ instead of $0.5^\circ$ to reduce the computational
cost). The two panels look very similar, showing that our secular
approach is reliable for systems that are far from a mean-motion
resonance.

Similar observations can be made for the ten extrasolar systems
considered here. In Fig.~\ref{fig:LK_megno}, the chaotic region
associated to a mean MEGNO value greater than eight with our
analytical approach, is indicated in white on the plot showing the
libration amplitude of $\omega_1$ (Fig.~\ref{fig:lidov-kozai}). Also,
more information on the extent of the chaotic zone for each system can
be found in the last column of Table~\ref{tab:res}. The chaotic region
around the stable LK islands is broad for half of the systems
(HD~11506, HD~12661, HD~169830, HD~207832, and HD~4732), moderate for
the HD~142, HD~15487, and HD~74156 systems, and not significant for
the HD~134987 and HD~164922 systems, given the integration timescale
and the grid of initial conditions considered. For the first category
of systems, long-term regular evolutions of the orbits are only
possible for low mutual inclinations and, for higher mutual
inclinations, in the LK region, while in the two other cases regular
evolutions are also observed at high mutual inclinations outside the
LK regions.

\section{Conclusions}
\label{sec:ccl}

In this work, we studied the possibility for ten RV-detected
exoplanetary systems to be in a 3D configuration. Using a secular
Hamiltonian approximation (expansion in eccentricities and
inclinations), we studied the secular dynamics of possible 3D
planetary configurations of the systems. In particular, we determined
ranges of orbital and mutual inclinations for which the system is in a
LK-resonant state. Our results were compared with n-body simulations,
showing the accuracy of the analytical approach up to very high
inclinations ($\sim 70^\circ-80^\circ$). We showed that all the
systems considered here might be in a LK-resonant state for a
sufficiently mutually inclined orbit. By means of the MEGNO chaos
indicator, we revealed the extent of the chaotic zone surrounding the
stability islands of the LK resonance. Long-term regular evolutions of
the orbits are possible i) at low mutual inclinations and ii) at high
mutual inclinations, preferentially in the LK region, due to the
significant extent of the chaotic zone in many systems.

It should be stressed that the present work excludes systems whose
inner planet is close to the star. For those systems, relativistic
effects have to be considered and we leave for future work how their
inclusion will influence the extent of the LK region.

\begin{acknowledgements}
The authors thank the anonymous referee for her or his critical review of
the first version of the manuscript and useful
suggestions. M.V. acknowledges financial support from the FRIA
fellowship (F.R.S.-FNRS). The work of A.R. is supported by a
F.R.S.-FNRS research fellowship. Computational resources have been
provided by the PTCI (Consortium des Équipements de Calcul Intensif
CECI), funded by the FNRS-FRFC, the Walloon Region, and the University
of Namur (Conventions No. 2.5020.11, GEQ U.G006.15, 1610468 et
RW/GEQ2016).
\end{acknowledgements}

\bibliographystyle{aa}
\bibliography{biblio}

\end{document}